\documentclass{ws-spin}
\usepackage{multicol}

\newcommand{\aplref}{9}
\newcommand{\japref}{19}
\newcommand{\phdref}{21}

\begin{document}

\markboth{Kuntal Roy}{Ultra-low-energy straintronics using multiferroic composites}

\title{ULTRA-LOW-ENERGY STRAINTRONICS USING MULTIFERROIC COMPOSITES}

\author{KUNTAL ROY\footnote{Current affiliation: School of Electrical and Computer Engineering, Purdue University, West Lafayette, IN 47907, USA.}}

\address{Electrical and computer Engineering Department\\
Virginia Commonwealth University, Richmond\\
VA 23284, USA\\
\email{royk@vcu.edu}
}

\maketitle


\begin{abstract}
This paper reviews the recent developments on building nanoelectronics for our future information processing paradigm using multiferroic composites. With appropriate choice of materials, when a tiny voltage of few tens of millivolts is applied across a multiferroic composite, i.e. a piezoelectric layer stain-coupled with a magnetostrictive layer, the piezoelectric layer gets strained and the generated stress in the magnetostrictive layer switches the magnetization direction between its two stable states. We particularly review the switching dynamics of magnetization and calculation of associated metrics like switching delay and energy dissipation. Such voltage-induced magnetization switching mechanism dissipates a minuscule amount of energy of only $\sim$1 attojoule in sub-nanosecond switching delay at room-temperature. The performance metrics for such non-volatile straintronic devices make them very attractive for building not only memory devices but also building logic, so that they can be deemed suitable for computational purposes. Hence, multiferroic straintronics has profound promise of contributing to Beyond Moore's law technology, i.e. of being possible replacement of conventional charge-based electronics, which is reaching its performance limit specifically due to excessive energy dissipation.
\end{abstract}

\keywords{Nanoelectronics; energy-efficient design; spintronics; straintronics; multiferroics.}

\begin{multicols}{2}

\section{Introduction}

The conventional charge-based electronics for more than past fifty years has a history of great success\cite{RefWorks:553}. However, the proven concept of enhancing the performance metrics by miniaturization\cite{moore65} of devices is approaching its fundamental limits\cite{RefWorks:126}; while there are issues due to process variation, basically the excessive energy dissipation in the devices limits the further improvement of transistor-based electronics\cite{RefWorks:211}. The Nanoelectronics Research Initiative (NRI)\cite{nri} at United States says ``Future generations of electronics will be based on new devices and circuit architectures, operating on physical principles that cannot be exploited by conventional transistors. NRI seeks the next device that will propel computing beyond the limitations of current technology.'' The challenge is to invent switching devices that dissipate miniscule amount of energy, e.g. $\sim$1 attojoule while maintaining sub-nanosecond switching delay. The performance metric switching delay is particularly important because if transistors were to switch at low switching speed there would not be excessive energy dissipation due to delay-energy trade-off. This is quite explicitly pointed out by NRI while seeking new switching devices over traditional transistors since several proposals could not meet the requirement of switching speed\cite{nri}.

Devices made of electron's spin as a state variable are switched by flipping spins without moving any charge in space and causing current flow. This eliminates the ohmic loss associated with switching, although some energy is still dissipated in flipping spins from one state to another overcoming the switching barrier between the states. It is widely believed that using ``spin'' as state variable is advantageous over the charge-based counterpart. Unfortunately, however, this advantage will be squandered if the method adopted to switch the spin is so energy-inefficient that the energy dissipated in the switching circuit far exceeds the energy dissipated inside the switch. Regrettably, this is often the case, e.g. switching spins with a magnetic field\cite{RefWorks:142,RefWorks:124} or with spin-transfer-torque mechanism\cite{RefWorks:7} using an external charge current\cite{roy11}. Hence, there is a need for inventing a switching mechanism to flip spins for the paradigm of spintronics to be technologically viable.

According to Brown's fundamental fine-ferromagnetic-particle theory\cite{RefWorks:490}, magnetic domain formation should be limited to very small dimensions ($\sim$100 nm) because of the competition between the magnetostatic energy and the quantum-mechanical {exchange energy}, causing nanomagnets to behave like single giant spins. These ``giant'' spins can beat superparamagnetic limit at room-temperature\cite{RefWorks:426}, which is crucial for general-purpose information processing. The minimum energy dissipated to switch such a single-domain nanomagnet (a collection of $M$ spins) can be only $\sim$$kTln(1/p)$, where $T$ is temperature and $p$ is error probability, since the exchange interaction between spins makes $M$ spins rotate together in unison like a giant \emph{classical} spin\cite{RefWorks:121,RefWorks:133}.  On the contrary, the minimum energy dissipated to switch a charge-based device like a transistor would be $\sim$$NkTln(1/p)$, where $N$ is the number of information carriers. This gives nanomagnets an \emph{inherent} advantage over the traditional transistors with regards to energy dissipation. 

In \emph{multiferroics}, different ferroic orders such as ferroelectric, ferromagnetic/ferrimagnetic, ferroelastic etc. coexist. For our discussion, we will assume the coexistence of ferroelectric and ferromagnetic orders to mean mutiferroism. Single-phase multiferroic materials are rare and moreover the magnetoelectric responses of those is very weak or occur only at low temperatures so their technological applications is not yet very promising\cite{RefWorks:558}. On the contrary, 2-phase multiferroic \emph{composites} consisting of magnetostrictive layers strain-coupled with piezoelectric layers\cite{RefWorks:558,Refworks:164,Refworks:165,RefWorks:562,RefWorks:328} do not have the bottlenecks as of their single-phase counterparts. Thus multiferroic composites are more promising for technological applications.

\vspace*{5mm}
\begin{figurehere}
\centerline{\psfig{file=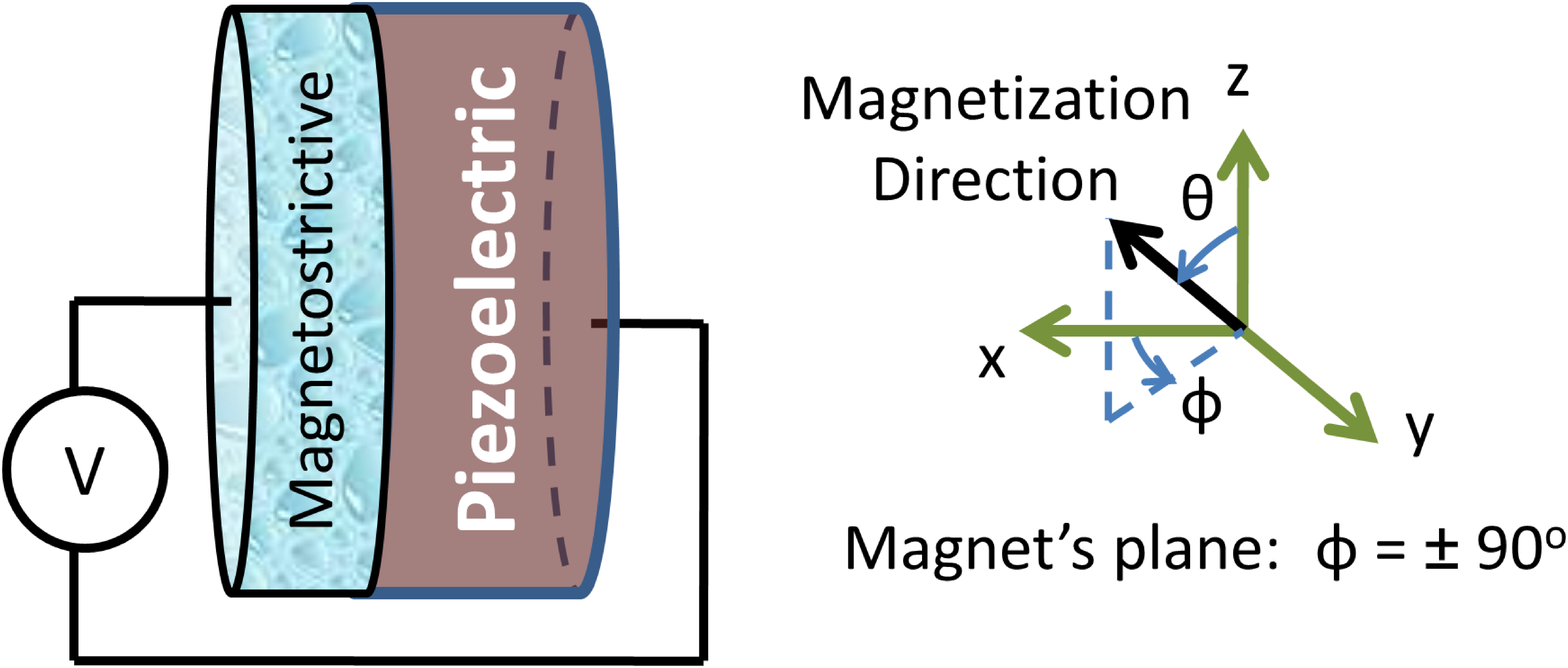,width=3.2in}}
\caption{A 2-phase multiferroic nanomagnet in the shape of 
an elliptical cylinder is stressed with an applied voltage via the $d_{31}$ coupling in the piezoelectric. The multiferroic is prevented from expanding or contracting along the in-plane hard axis ($y$-axis), so that a uniaxial stress is generated along the easy axis ($z$-axis). (Reprinted with permission from Ref.~\japref. Copyright 2012, AIP Publishing LLC.)}
\label{fig:multiferroic} 
\end{figurehere}

The magnetization of the shape-anisotropic single-domain magnetostrictive nanomagnet in multiferroic composites can be switched in less than 1 nanosecond while dissipating only $\sim$1 attojoule\cite{roy11,roy11_6}. Hence such devices have emerged as potential candidates as storage and switching elements for our future non-volatile memory and logic systems. Particularly due to high switching speed while simultaneously being highly energy efficient, this has lead to logic proposals incorporating such systems\cite{RefWorks:154,roy_phd,fasha11,nano_edi}. The magnetization of the nanomagnet has two stable states (mutually anti-parallel) along the easy axis encoding the binary bits 0 and 1. The magnetization can be switched from one stable state to the other when a tiny voltage of few tens of millivolts is applied across the piezoelectric layer while constraining it from expanding or contracting along its in-plane hard-axis (see Fig.~\ref{fig:multiferroic}). The applied voltage produces a strain in the piezoelectric layer, which is then transferred to the magnetostrictive layer. This in turn generates a uniaxial stress in the magnetostrictive nanomagnet along its easy-axis and rotates the magnetization towards the in-plane hard axis as long as the product of the stress and the magnetostrictive coefficient is {\it negative}. It is assumed by convention that a tensile stress is positive and a compressive stress is negative. There have been experimental efforts to demonstrate such electric-field induced magnetization rotation\cite{RefWorks:550,RefWorks:551,RefWorks:559,RefWorks:167,RefWorks:609}.

The rest of the paper is organized as follows. In Section~\ref{sec:model}, we describe the model for such multiferroic straintronic devices. The Landau-Lifshitz-Gilbert (LLG) equation is solved to track the magnetization dynamics and to calculate the associated performance metrics like switching delay and energy dissipation. Also, we present the model for determining magnetization dynamics in a circuit made of multiple multiferroic devices. Section~\ref{sec:results} presents the simulation results for both a single memory device and a circuit of multiple multiferroic devices. Finally, Section~\ref{sec:conclusions} summarizes this review and provides the outlook of the multiferroic straintronic devices on building nanoelectronics for our future information processing paradigm.

\section{\label{sec:model}Model}

In this Section, we will first review the model of a single multiferroic device. The emphasis would be on magnetization dynamics of the magnetostrictive nanomagnet in a multiferroic composite by solving Landau-Lifshitz-Gilbert (LLG) equation\cite{RefWorks:162,RefWorks:161}. We will review the model incorporating room-temperature thermal fluctuations using \emph{stochastic} LLG equation\cite{RefWorks:186,roy11_6}. Then we describe how the same model can be used for \emph{unidirectional} signal propagation in a chain of multiferroic devices\cite{roy_phd,fasha11}. How we can build logic gates for computational purposes is described too in the end.

\subsection{Single Multiferroic Device}

We consider a single isolated nanomagnet in the shape of an elliptical cylinder with its elliptical cross section lying in the $y$-$z$ plane; the major axis is aligned along the $z$-direction and minor axis along the $y$-direction (see Fig.~\ref{fig:multiferroic}). The dimensions of the major axis, the minor axis, and the thickness are $a$, $b$, and $l$, respectively. So the magnet's volume is $\Omega=(\pi/4)abl$. The $z$-axis is the easy axis, the $y$-axis is the in-plane hard axis and the $x$-axis is the out-of-plane hard axis. Since $l \ll b$, the out-of-plane hard axis is much harder than the in-plane hard axis. Let $\theta(t)$ be the polar angle and $\phi(t)$ the azimuthal angle of the magnetization vector in standard spherical coordinate system. Note that when $\phi$ = $\pm90^{\circ}$, the magnetization vector lies in the plane of the nanomagnet. Any deviation from $\phi$ = $\pm90^{\circ}$ corresponds to out-of-plane excursion.

We can write the total energy of the magnetostrictive single-domain nanomagnet when it is subjected to uniaxial stress along the easy axis (major axis of the ellipse) as the sum of the uniaxial shape anisotropy energy and the uniaxial stress anisotropy energy\cite{RefWorks:157}. We assume that the magnetostrictive layer is polycrystalline, so that we ignore the magnetocrystalline energy. The uniaxial shape anisotropy energy at an instant of time $t$ is given by\cite{RefWorks:157} 
\begin{align}
E_{SHA}(t) &= E_{SHA}(\theta(t),\phi(t)) \nonumber\\
					 &= (\mu_0/2) M_s^2 \Omega N_d(\theta(t),\phi(t))
\label{eq:shape_anisotropy}
\end{align}
where $M_s$ is the saturation magnetization and the demagnetization factor $N_d(t)$ is expressed as\cite{RefWorks:157} 
\begin{multline}
N_d(t) = N_d(\theta(t),\phi(t)) = N_{d-xx} sin^2\theta(t) \, cos^2\phi(t) \\ + N_{d-yy} sin^2\theta(t) \ sin^2\phi(t) + N_{d-zz} cos^2\theta(t) 
\end{multline}
with $N_{d-mm}$ being the m$^{th}$ (m=x,y,z) component of the demagnetization factor\cite{RefWorks:402}. Note that these factors depend on the dimensions of the nanomagnet and not on the material properties. The dimensions of the nanomagnet is chosen as $a$ = 100 nm, $b$ = 90 nm and $l$ = 6 nm, which ensures that the nanomagnet has a single ferromagnetic domain\cite{RefWorks:133}. These dimensions alongwith the material parameter saturation magnetization $M_s$ determine the shape anisotropy energy barrier, which separates the two stable states of the nanomagnet. The in-plane energy barrier $E_b$ ($\phi = \pm 90^{\circ}$, see Fig.~\ref{fig:multiferroic}), which is the \emph{lowest} difference between the shape anisotropy energies when $\theta = 90^{\circ}$ and $\theta = 0^{\circ}, 180^{\circ}$ determines the \emph{static} error probability of spontaneous magnetization reversal due to thermal fluctuations. According to Boltzmann distribution, this probability is $\exp \left [-E_b/kT \right ]$. This probability should be low enough for technological application purposes. With the dimensions and material chosen, $E_b$ = 44 $kT$ at room temperature, so that the static error probability at room temperature is $e^{-44}$. Note that the \emph{dynamic} error probability on the other hand signifies the switching error probability when magnetization fails to flip from one state to another during switching events. Usually the \emph{dynamic} error probability can be much higher than that of \emph{static} error probability. So we must put emphasis on this \emph{dynamic} error probability to meet the demand of technological viability.

The stress anisotropy energy is given by\cite{RefWorks:557,RefWorks:157} 
\begin{align}
E_{STA}(t) &= E_{STA}(\theta(t),\sigma(t)) \nonumber\\
					 &= - (3/2) \lambda_s \sigma(t) \Omega \, cos^2\theta(t)
\label{eq:stress_anisotropy}
\end{align}
where $(3/2) \lambda_s$ is the magnetostriction coefficient of the magnetostrictive nanomagnet and $\sigma(t)$ is the stress generated in it by an external voltage. A positive $\lambda_s \sigma(t)$ product will favor alignment of the magnetization along the major axis ($z$-axis), while a negative $\lambda_s \sigma(t)$ product will favor alignment along the minor axis ($y$-axis), because that will minimize $E_{STA}(t)$. We will use the following convention that a compressive stress is negative and tensile stress is positive. Therefore, in a material like Terfenol-D that has positive $\lambda_s$, a compressive stress will favor alignment along the minor axis, and tensile along the major axis. The situation will be exactly opposite with nickel and cobalt that have negative $\lambda_s$.

At any instant of time $t$, the total energy of the nanomagnet can be expressed as\cite{roy11_6}
\begin{align}
E(t) &= E(\theta(t),\phi(t),\sigma(t)) \nonumber\\
		 &= B(\phi(t),\sigma(t)) sin^2\theta(t) + C(t)
\end{align}
where 
\begin{subequations}
\begin{align}
B(t) &= B(\phi(t),\sigma(t)) = B_0(\phi(t)) + B_{stress}(\sigma(t)) \displaybreak[3]\\
B_0(t) &= B_0(\phi(t)) = (\mu_0/2) \, M_s^2 \Omega \lbrack N_{d-xx} cos^2\phi(t) \nonumber \\
			& \qquad + N_{d-yy} sin^2\phi(t) - N_{d-zz}\rbrack \displaybreak[3]\\
B_{stress}(t) &= B_{stress}(\sigma(t)) = (3/2) \lambda_s \sigma(t) \Omega \displaybreak[3]\\
C(t) &= C(\sigma(t)) = (\mu_0/2) M_s^2 \Omega N_{d-zz} \nonumber\\
			& \hspace*{2.5cm} - (3/2) \lambda_s \sigma(t) \Omega. \displaybreak[3]
\end{align}
\end{subequations}

The magnetization \textbf{M}(t) of the nanomagnet has a constant magnitude at any given temperature but a variable direction, so that we can represent it by the vector of unit norm $\mathbf{n_m}(t) =\mathbf{M}(t)/|\mathbf{M}| = \mathbf{\hat{e}_r}$ where $\mathbf{\hat{e}_r}$ is the unit vector in the radial direction in spherical coordinate system represented by ($r$,$\theta$,$\phi$). The other two unit vectors in the spherical coordinate system are denoted by $\mathbf{\hat{e}_\theta}$ and $\mathbf{\hat{e}_\phi}$ for $\theta$ and $\phi$ rotations, respectively. The torque acting on the magnetization per unit volume due to shape and stress anisotropy is\cite{roy11_6}
\begin{align}
\mathbf{T_E} (t) &= - \mathbf{n_m}(t) \times \nabla E(\theta(t),\phi(t),\sigma(t)) \nonumber\\
&= - 2 B(\phi(t),\sigma(t)) sin\theta(t) cos\theta(t) \,\mathbf{\hat{e}_\phi} \nonumber\\
& \qquad - B_{0e}(\phi(t)) sin\theta (t) \,\mathbf{\hat{e}_\theta}, 							 
\label{eq:stress_torque}
\end{align}
\noindent
where 
\begin{align}
B_{0e}(t) &= B_{0e}(\phi(t)) \nonumber\\
					&= (\mu_0/2) \, M_s^2 \Omega (N_{d-xx}-N_{d-yy}) sin(2\phi(t)). 
\label{eq:B0e}
\end{align}

The effect of room-temperature random thermal fluctuations is incorporated via a random magnetic field $\mathbf{h}(t)$, which is expressed as 
\begin{equation}
\mathbf{h}(t) = h_x(t)\mathbf{\hat{e}_x} + h_y(t)\mathbf{\hat{e}_y} + h_z(t)\mathbf{\hat{e}_z}
\end{equation}
\noindent
where $h_i(t)$ (i=x,y,z) are the three components of the random thermal field in Cartesian coordinates. We assume the properties of the random field $\mathbf{h}(t)$ as described in Ref.~\refcite{RefWorks:186}. The random thermal field can be written as\cite{roy11_6}
\begin{equation}
h_i(t) = \sqrt{\frac{2 \alpha kT}{|\gamma| M_V \Delta t}} \; G_{(0,1)}(t) \quad (i=x,y,z)
\label{eq:ht}
\end{equation}
\noindent
where $\alpha$ is the dimensionless phenomenological Gilbert damping parameter, $\gamma = 2\mu_B \mu_0/\hbar$ 
is the gyromagnetic ratio for electrons and is equal to $2.21\times 10^5$ (rad.m).(A.s)$^{-1}$, $\mu_B$ is the Bohr magneton, $M_V= \mu_0 M_s \Omega$, and $1/\Delta t$ is proportional to the attempt frequency of the thermal field, $\Delta t$ is the simulation time-step used, and the quantity $G_{(0,1)}(t)$ is a Gaussian distribution with zero mean and unit variance\cite{RefWorks:388}. 

The thermal torque can be written as\cite{roy11_6}
\begin{equation}
\mathbf{T_{TH}}(t) = M_V\,\mathbf{n_m}(t) \times \mathbf{h}(t) = P_\theta(t)\,\mathbf{\hat{e}_\phi} - P_\phi(t)\,\mathbf{\hat{e}_\theta}
\end{equation}
\noindent
where
\begin{eqnarray}
P_\theta(t) &=& M_V\lbrack h_x(t)\,cos\theta(t)\,cos\phi(t) \nonumber\\ 
						&& + h_y(t)\,cos\theta(t)sin\phi(t) - h_z(t)\,sin\theta(t) \rbrack,\nonumber\\
						&&\\
P_\phi(t) &=& M_V \lbrack h_y(t)\,cos\phi(t) -h_x(t)\,sin\phi(t)\rbrack.
\label{eq:thermal_parts}
\end{eqnarray}
\noindent

The magnetization dynamics under the action of the torques $\mathbf{T_{E}}(t)$ and 
$\mathbf{T_{TH}}(t)$ is described by the stochastic Landau-Lifshitz-Gilbert (LLG) equation as follows.
\begin{multline}
\cfrac{d\mathbf{n_m}(t)}{dt} - \alpha \left(\mathbf{n_m}(t) \times \cfrac{d\mathbf{n_m}(t)}{dt} \right)\\
 = -\cfrac{|\gamma|}{M_V} \left\lbrack \mathbf{T_E}(t) +  \mathbf{T_{TH}}(t)\right\rbrack.
\end{multline}

After solving the above equation analytically, we get the following coupled equations of magnetization dynamics for $\theta(t)$ and $\phi(t)$\cite{roy11_6}.
\begin{multline}
\left(1+\alpha^2 \right) \cfrac{d\theta(t)}{dt} = \frac{|\gamma|}{M_V} \lbrack  B_{0e}(\phi(t)) sin\theta(t) \\ 
				- 2\alpha B(\phi(t),\sigma(t)) sin\theta (t)cos\theta (t) \\
				+ \left(\alpha P_\theta(t) + P_\phi (t) \right) \rbrack,
 \label{eq:theta_dynamics}
\end{multline}
\begin{multline}
\left(1+\alpha^2 \right) \cfrac{d \phi(t)}{dt} = \frac{|\gamma|}{M_V} \lbrack \alpha B_{0e}(\phi(t)) \\
					+ 2 B(\phi(t),\sigma(t)) cos\theta(t) \\ 
					- {[sin\theta(t)]^{-1}} \left(P_\theta (t) - \alpha P_\phi (t) \right) \rbrack \\
	(sin\theta \neq 0).
\label{eq:phi_dynamics}
\end{multline}
We need to solve the above two coupled equations numerically to track the trajectory of magnetization over time, in the presence of thermal fluctuations.

When $\sin\theta=0$ ($\theta=0^\circ$ or $\theta=180^\circ$), i.e. when the magnetization direction is {\it exactly} along the easy axis, the torque on the magnetization vector given by Eq.~\eqref{eq:stress_torque} becomes zero. That is why only thermal fluctuations can budge the magnetization vector from the easy axis. Consider the situation when $\theta=180^\circ$. From Eqs.~\eqref{eq:theta_dynamics} and~\eqref{eq:phi_dynamics}, we get\cite{roy11_6}
\begin{equation}
\phi(t) = tan^{-1} \left( \frac{\alpha h_y(t) + h_x(t)}{h_y(t) - \alpha h_x(t)} \right),
\label{eq:phi_t_thermal}
\end{equation}
\begin{equation}
\cfrac{d\theta(t)}{dt} =  \frac{-|\gamma| (h_x^2(t) + h_y^2(t))}{\sqrt{(h_y(t)-\alpha h_x(t))^2 + (\alpha h_y(t) + h_x(t))^2}}.
\label{eq:theta_t_thermal}
\end{equation}
\noindent
We can see from the Eq.~\eqref{eq:theta_t_thermal} clearly that thermal torque can deflect the magnetization from the easy axis since the time rate of change of $\theta(t)$ would be non-zero in the presence of the thermal fluctuations. Note that $d\theta(t)/dt$ does not depend on the component of the random thermal field along the $z$-axis, i.e. $h_z(t)$, which is a consequence of having $z$-axis as the easy axis of the nanomagnet. However, once the magnetization direction is even slightly deflected from the easy axis, all three components of the random thermal field along the $x$-, $y$-, and $z$-direction would come into play.

When no stress is applied on the magnetostrictive nanomagnet, magnetization would just fluctuate around an easy axis provided that the shape anisotropy energy barrier is enough high to prevent spontaneous reversal of magnetization from one state to another in a short period of time. We can solve the Eqs.~\eqref{eq:theta_dynamics} and~\eqref{eq:phi_dynamics} while setting $B_{stress}$ = 0 to track the dynamics of magnetization due to thermal fluctuations. So this will yield the distribution of the magnetization vector's initial orientation when stress is turned on. Since the most probably value of magnetization is along easy axis, the $\theta$-distribution is Boltzmann peaked at $\theta$ = 0$^{\circ}$ or 180$^{\circ}$, while the $\phi$-distribution is Gaussian peaked at $\phi = \pm 90^{\circ}$ because these positions are minimum energy positions\cite{roy11_5}. 

Stress is ineffective when $\theta$ is around 0$^{\circ}$ or 180$^{\circ}$, i.e. when magnetization is around the easy axis (mathematically, note that the expression in Eq.~\eqref{eq:stress_torque} is proportional to $sin\,\theta$, which is zero when $\theta$ is equal to 0$^{\circ}$ or 180$^{\circ}$). Hence, we would get a long tail in the switching delay distribution should magnetization starts very near from easy axis. When we start out from $\theta = 0^{\circ}, 180^{\circ}$, we have to wait a while; thermal fluctuations may help here while getting started but random thermal kicks may also cause magnetization to traverse towards the opposite direction than the intended dirction of switching. Thus, switching trajectories initiating from near easy axis may be very slow. We do need to worry about the switching delay \emph{tail} more than \emph{mean} switching delay since the extent of tail will set the requirement of pulse width of stress for switching to take place with sufficiently high probability.

In order to eliminate the long tail in the switching delay distribution, we can apply a static bias field that will shift the peak of $\theta_{initial}$ distribution away from the easy axis, so that the most probable starting orientation will no longer be the easy axis\cite{roy11_6}. This field is applied along the out-of-plane hard axis (+$x$-direction) and the potential energy due to the applied magnetic field can be expressed as
\begin{equation}
E_{mag}(\theta(t),\phi(t)) = - M_V H\, sin\theta(t)\,cos\phi(t),
\end{equation}
where $H$ is the magnitude of magnetic field. The torque generated due to this field is 
\begin{equation}
\mathbf{T_M} (t) = - \mathbf{n_m}(t) \times \nabla E_{mag}(\theta(t),\phi(t)). 
\end{equation}
We assume that a permanent magnetic sheet will be employed to produce the bias field and thus will not require any additional energy dissipation to be generated. The presence of this field will modify Eqs.~\eqref{eq:theta_dynamics} and~\eqref{eq:phi_dynamics} for magnetization dynamics to\cite{roy11_6} 
\begin{multline}
\left(1+\alpha^2 \right) \cfrac{d\theta(t)}{dt} = \frac{|\gamma|}{M_V} \times\\
	 \lbrack B_{0e}(\phi(t)) sin\theta(t) - 2\alpha B(\phi(t),\sigma(t)) sin\theta (t)cos\theta (t) \\
	 + \alpha M_V H\, cos\theta(t)\,cos\phi(t) - M_V H\, sin\phi(t) \\
 	 + \left(\alpha P_\theta (t) + P_\phi (t) \right) \rbrack,
 \label{eq:theta_dynamics_mag}
\end{multline}
\begin{multline}
\left(1+\alpha^2 \right) \cfrac{d \phi(t)}{dt} = \frac{|\gamma|}{M_V} \times \\
		\lbrack \alpha B_{0e}(\phi(t)) + 2 B(\phi(t),\sigma(t)) cos\theta(t) \\
   - {[sin\theta(t)]^{-1}} \left(M_V H\, cos\theta(t)\,cos\phi(t) + \alpha M_V H\, sin\phi(t)   \right) \\
   - {[sin\theta(t)]^{-1}} \left(P_\theta (t) - \alpha P_\phi (t) \right) \rbrack \quad	(sin\theta \neq 0).
\label{eq:phi_dynamics_mag}
\end{multline}
\noindent

Note that the bias field makes the potential energy profile of the nanomagnet asymmetric in $\phi$-space and the energy minimum gets shifted from $\phi_{min}=\pm90^\circ$ (the plane of the nanomagnet) to
\begin{equation}
\phi_{min} = cos^{-1}\left\lbrack \frac{H}{M_s (N_{d-xx} - N_{d-yy})} \right\rbrack.
\end{equation}
\noindent
However, the potential profile will remain symmetric in $\theta$-space, with $\theta=0^\circ$ and $\theta=180^\circ$ remaining as the minimum energy locations. A bias magnetic field of flux density 40 mT applied perpendicular to the plane of the magnet would make $\phi_{min} \simeq \pm87^\circ$ deflecting the magnetization vector $\sim$$3^{\circ}$ from the magnet's plane. Application of the bias magnetic field will also affect the in-plane shape anisotropy energy barrier $E_b$ as it gets reduced from 44 $kT$ to 36 $kT$ at room temperature.

We consider both the energy dissipated internally in the nanomagnet due to Gilbert damping (termed as $E_d$) and the energy dissipated in the switching circuit while applying voltage across the multiferroic structure generating stress on the nanomagnet (termed as `$CV^2$' dissipation, where $C$ and $V$ denote the capacitance of the piezoelectric layer and the applied voltage, respectively). If the voltage is turned on or off abruptly then the energy dissipated during either turn on or turn off is $(1/2)CV^2$, however, if the ramp rate is finite, the energy dissipated can be significantly reduced\cite{roy11_2}. The internal energy dissipation $E_d$, is given by the expression $\int_0^{\tau}P_d(t) dt$, where $\tau$ is the switching delay and $P_d(t)$ is the power dissipated during switching given as\cite{roy11_6}
\begin{equation}
P_d(t) = \frac{\alpha \, |\gamma|}{(1+\alpha^2) M_V} \, \left| \mathbf{T_E}(t) + \mathbf{T_M}(t)\right|^2 .
\label{eq:power_dissipation}
\end{equation}
\noindent
 We sum up the power $P_d(t)$ dissipated during the entire switching period to get the corresponding energy dissipation $E_d$ and add that to the `$CV^2$' dissipation in the switching circuit to find the total dissipation $E_{total}$. There is no net dissipation due to random thermal torque, however, it affects $E_d$ since it raises the critical stress needed to switch with $\sim$100\% probability and it also affects the stress needed to switch with a given probability.

\subsection{Array of Multiferroic Devies}

Here we will use the same model as derived for a single multiferroic device and see how \emph{unidirectional} flow of signal is possible in a horizontal chain of multiferroic devices\cite{RefWorks:154,roy_phd,fasha11} using dipole coupling between nanomagnets and Bennett clocking mechanism\cite{RefWorks:144}. 

The dipole coupling between two magnetic moments $\mathbf{M1}$ and $\mathbf{M2}$ separated by a distance vector $\mathbf{R}$ can be expressed as\cite{RefWorks:158}:
\begin{equation}
	E_{dipole} = \frac{1}{4\pi \mu_0 R^3} \left\lbrack (\mathbf{M_1.M_2}) - \frac{3}{R^2} (\mathbf{M_1.R}) (\mathbf{M_2.R})\right\rbrack.
	\label{eq:dipole_coupling}
\end{equation}

\begin{figurehere}
\centerline{\psfig{file=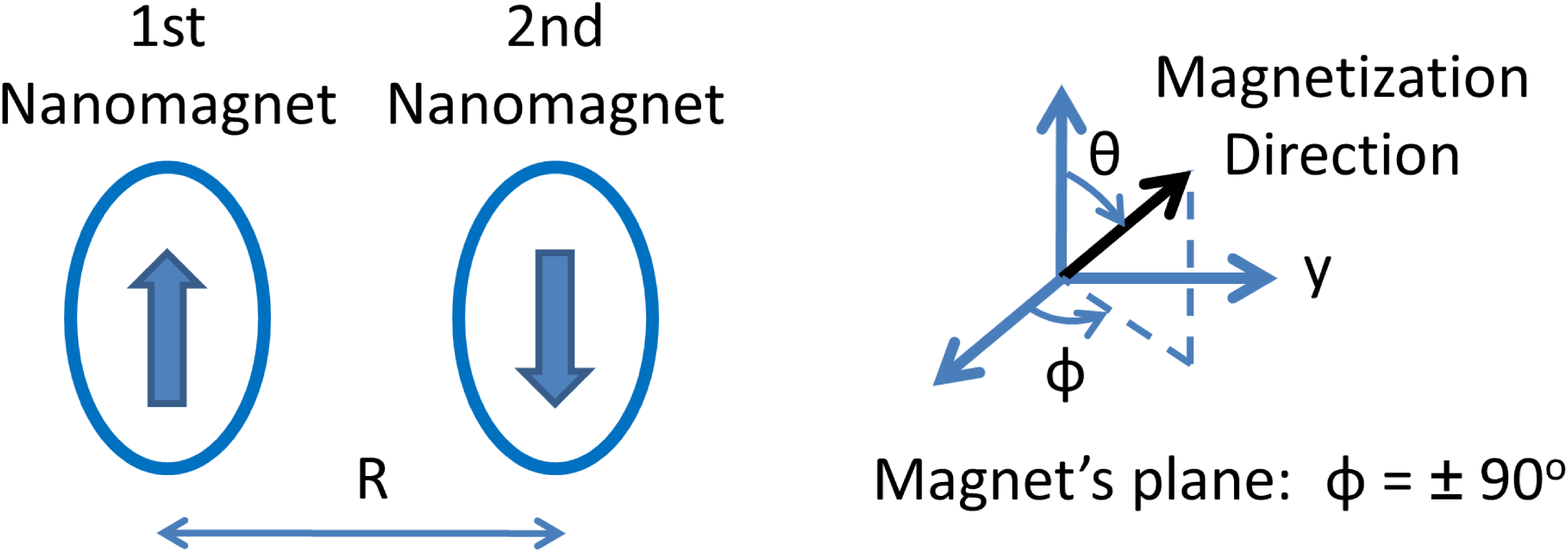,width=3.2in}}
\caption{\label{fig:dipole_coupling}Dipole coupling between two magnetic moments. (Reprinted with permission from Ref.~\phdref. Copyright 2012, Kuntal Roy.)}
\end{figurehere}

In standard spherical coordinate system (see Fig.~\ref{fig:dipole_coupling}), the expression of dipole coupling can be formulated as 
\begin{align}
	E_{dipole} &= \cfrac{\mu_0}{4\pi R^3} \, M_s^2 \Omega^2 \lbrack cos\theta_1 cos\theta_2 \nonumber\\ 
						 &\;+ sin \theta_1 sin \theta_2 (cos \phi_1 cos\phi_2 - 2 sin\phi_1 sin\phi_2) \rbrack 
	\label{eq:dipole_coupling_spherical}
\end{align}
\noindent
where $|\mathbf{M_1}|=|\mathbf{M_2}|=\mu_0 M_s \Omega$, $\Omega$ is the volume of the nanomagnets, $M_s$ is the saturation magnetization, and $\mathbf{R}=R\,\mathbf{\hat{e}_y}$.

Note that dipole coupling is bi-directional, i.e. $E_{dipole}=E_{dipole,1}=E_{dipole,2}$. Because of the dipole coupling between the magnetizations of the nanomagnets, the potential profiles of both the nanomagnets are tilted and the ground state of the magnetizations are antiferromagnetically coupled as depicted in the Fig.~\ref{fig:dipole_coupling}.

\begin{figure*}
\begin{center}
\psfig{file=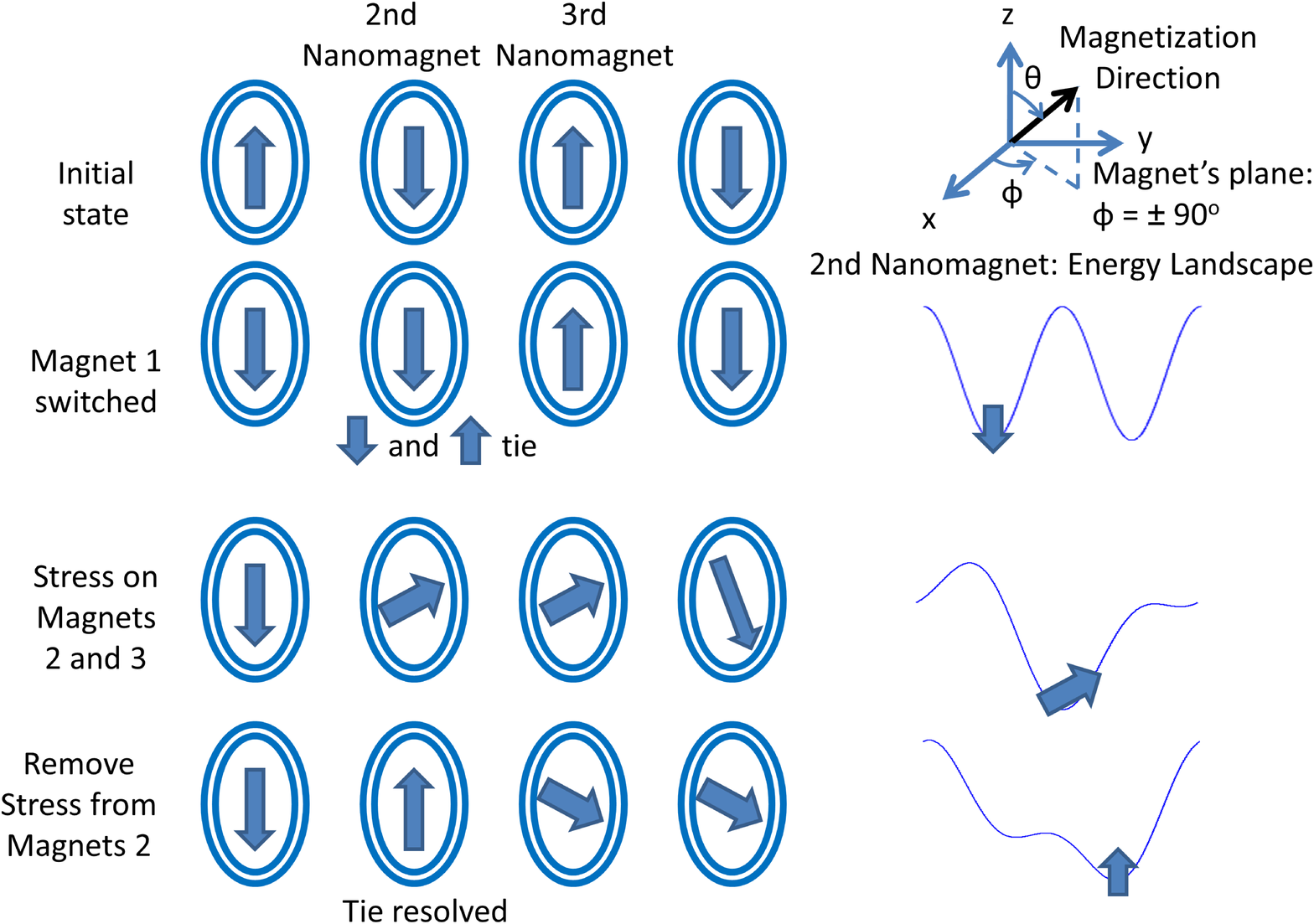,width=6in}
\end{center}
\caption{\label{fig:bennet_clocking_problem_solved_3D} Imposing the unidirectionality in time to propagate a logic bit through a chain of nanomagnets. The 2nd and 3rd nanomagnets are stressed to align their magnetizations along the hard axis and then stress is removed/reversed on the 2nd nanomagnet to relax its magnetization along the desired state. (Reprinted with permission from Ref.~\phdref. Copyright 2012, Kuntal Roy.)}
\end{figure*}

If we somehow change the magnetization direction of one nanomagnet, the magnetization of the other nanomagnet would \emph{not} automatically change its direction to assume an antiferromagnetic order. It is because of the reason that there is a barrier separating two magnetization states. It's true that antiferromagnetic order is the ground state, however, during operation of devices, we must remove the barrier and then again restore it to make sure that antiferromagnetic order is maintained. Magnetization may come to antiferromagnetic order after a very long time depending on the barrier height but the operation of devices cannot be dependent on that.

In general, we need to propagate a logic bit \emph{unidirectionally} along a chain of nanomagnets. It requires a clock signal to periodically reset the magnetization direction of each nanomagnet. If a global magnetic field is utilized for such a purpose, it would not allow pipelining of data, and magnetization of every nanomagnet must be maintained along hard axis until a bit propagates. It needs an energy minima along hard axis, which can be introduced by biaxial anisotropy\cite{RefWorks:140}, but thermal fluctuations would produce a large bit error probability\cite{RefWorks:160}. Using a local magnetic field eliminates the problems of using global magnetic field, but it is difficult to maintain a magnetic field locally within a dimension of $\sim$100 nm. Furthermore, generating magnetic field is highly energy consuming. We can use electric-field operated (since electric-field can be maintained locally) multiferroic devices to propagate signals in a chain of nanomagnets\cite{RefWorks:154} using so-called Bennett clocking mechanism, termed in the name of Bennett\cite{RefWorks:144}. Ref.~\refcite{RefWorks:154} performed the steady-state analysis, while Refs.~\refcite{roy_phd,fasha11} solved the magnetization dynamics using the same model as for a single multiferroic device to show that the switching may take place in sub-nanosecond\cite{roy11,roy11_2,roy11_6}, which is crucial for building nanomagnetic logic\cite{nano_edi}.

Fig.~\ref{fig:bennet_clocking_problem_solved_3D} depicts the issue (and also solution) behind Bennett clocking in a chain of nanomagnet. First of all, it needs to be emphasized that dipole coupling is bi-directional. So the 2nd nanomagnet experiences dipole coupling effect from both of its neighbors, i.e. 1st and 3rd nanomagnets. Note that we are considering only nearest neighbor interaction, since dipole coupling reduces drastically with distance [see Eq.~\eqref{eq:dipole_coupling}]. Thus, if the 1st nanomagnet is switched, the 2nd nanomagnet finds itself in a locked condition as the 1st nanomagnet is telling it to go \emph{up}, while the 3rd nanomagnet is telling it to go \emph{down}. Therefore, it remains on its previous position and thus the change in information on the 1st nanomagnet cannot be propagated through the chain of nanomagnets.

To prevent this lockjam, we need to impose the \emph{unidirectionality in time} as shown in the Fig.~\ref{fig:bennet_clocking_problem_solved_3D}. Both the 2nd and 3rd nanomagnets are stressed to get them aligned to their hard axes (note the third row in Fig.~\ref{fig:bennet_clocking_problem_solved_3D}) and then stress is removed/reversed on the 2nd nanomagnet (note the fourth row in Fig.~\ref{fig:bennet_clocking_problem_solved_3D}) to relax its magnetization towards the desired state. In this way, subsequently applying stress on the nanomagnets and then releasing/reversing the stress, we can propagate a logic bit \emph{unidirectionally} along a chain of nanomagnets. The slight deflection in the magnetization of the 4th nanomagnet in the third row of Fig.~\ref{fig:bennet_clocking_problem_solved_3D} is due to dipole coupling, while in the fourth row, the magnetization of 4th nanomagnet is aligned along its hard axis because of applied stress on it. A 3-phase clock would be sufficient to propagate a signal along the chain of nanomagnet. Note that we are explaining the operation with two-dimensional in-plane potential landscapes of the nanomagnets (assuming azimuthal angle $\phi=\pm 90^\circ$), but solution of the full three-dimensional dynamics is necessary since the out-of-plane excursion of magnetization has immense influence in shaping the magnetization dynamics\cite{roy11_5} and reducing the switching delay by a couple of orders in magnitude to sub-nanosecond\cite{roy11,roy11_2,roy11_6}.

The first row in the Fig.~\ref{fig:bennet_clocking_problem_solved_3D} shows a chain of four nanomagnets and we intend to switch the 2nd nanomagnet successfully in its desired direction once the 1st nanomagnet is switched as depicted in the second row of Fig.~\ref{fig:bennet_clocking_problem_solved_3D}. We will use subscripts 1-4 to denote the parameters and metrics for the corresponding nanomagnets. The dipole coupling energy on the 2nd nanomagnet due to 1st and 3rd nanomagnets can be written following the similar prescription given in the Eq.~\eqref{eq:dipole_coupling_spherical} as\cite{roy_phd}
\begin{align}
	E_{dipole,2} &= \cfrac{\mu_0}{4\pi R^3} \, M_s^2 \Omega^2 \lbrack cos\theta_2 cos\theta_1  + cos\theta_2 cos\theta_3 \nonumber\\
	& \quad + sin \theta_1 sin \theta_2 (cos \phi_1 cos\phi_2 - 2 sin\phi_1 sin\phi_2) \nonumber\\
	& \quad + sin \theta_3 sin \theta_2 (cos \phi_3 cos\phi_2 - 2 sin\phi_3 sin\phi_2) \rbrack.
\end{align}
The torque acting on the 2nd nanomagnet due to dipole coupling\cite{roy_phd}
\begin{align}
\mathbf{T_{dipole,2}} (t) &= - \mathbf{n_m}(t) \times \nabla E_{dipole,2} \nonumber\\
								 &= - \cfrac{\partial E_{dipole,2}}{\partial \theta_2} \,\mathbf{\hat{e}_\phi} +  \cfrac{1}{sin \theta_2}\,\cfrac{\partial E_{dipole,2}}{\partial \phi_2} \,\mathbf{\hat{e}_\theta} \nonumber\\
								 &= - T_{dipole,\phi_2} \mathbf{\hat{e}_\phi} + T_{dipole,\theta_2} \mathbf{\hat{e}_\theta},
\end{align}
\noindent
where

\begin{align}
	T_{dipole,\phi_2} &= \cfrac{\partial E_{dipole,2}}{\partial \theta_2} \nonumber\\
										&= \cfrac{\mu_0}{4\pi R^3} \, M_s^2 \Omega^2 \lbrack -sin\theta_2 cos\theta_1  - sin\theta_2 cos\theta_3 \nonumber\\
										&\quad+ sin \theta_1 cos \theta_2 (cos \phi_1 cos\phi_2 - 2 sin\phi_1 sin\phi_2) \nonumber\\
										&\quad+ sin \theta_3 cos \theta_2 (cos \phi_3 cos\phi_2 - 2 sin\phi_3 sin\phi_2) \rbrack,
\end{align}
\noindent
and
\begin{align}
	T_{dipole,\theta_2} &= \cfrac{1}{sin \theta_2}\,\cfrac{\partial E_{dipole,2}}{\partial \phi_2} \nonumber\\
											&= -\cfrac{\mu_0}{4\pi R^3} \, M_s^2 \Omega^2  \times \nonumber\\
											&\quad \lbrack sin \theta_1 (cos \phi_1 sin\phi_2 + 2 sin\phi_1 cos\phi_2) \nonumber\\
											&\quad\; + sin \theta_3 (cos \phi_3 sin\phi_2 + 2 sin\phi_3 cos\phi_2) \rbrack.
\end{align}
The torque acting on the 2nd nanomagnet due to shape and stress anisotropy can be derived similarly following the Eq.~\eqref{eq:stress_torque} as\cite{roy_phd}
\begin{align}
\mathbf{T_{E,2}} (t) &= - 2 B_2(\phi_2(t)) sin\theta_2(t) cos\theta_2(t) \,\mathbf{\hat{e}_\phi} \nonumber\\
										 & \quad  - B_{0 e2}(\phi_2(t)) sin\theta_2(t) \,\mathbf{\hat{e}_\theta},
\label{eq:stress_torque_bennett_2nd}
\end{align}
\noindent
where
\begin{subequations}
\begin{eqnarray}
B_2(\phi_2(t)) &=& \cfrac{\mu_0}{2} \, M_s^2 \Omega \lbrack N_{d-xx} cos^2\phi_2(t) \nonumber\\
							&& \quad  + N_{d-yy} sin^2\phi_2(t) - N_{d-zz}\rbrack \nonumber\\
							&& \qquad + (3/2) \, \lambda_s \sigma_2 \Omega,\\
B_{0e2}(\phi_2(t)) &=& \cfrac{\mu_0}{2} \, M_s^2 \Omega (N_{d-xx}-N_{d-yy}) sin(2\phi_2(t)). \nonumber\\
\end{eqnarray}
\end{subequations}

After solving the Landau-Lifshitz-Gilbert (LLG) equation considering the dipole coupling term in a very similar way as done for a single multiferroic device, we get the coupled dynamics between the polar angle $\theta_2$ and azimuthal angle $\phi_2$ for the 2nd nanomagnet as\cite{roy_phd}

\begin{align}
\left(1+\alpha^2 \right) \cfrac{d\theta_2(t)}{dt} &= \cfrac{|\gamma|}{M_V} \lbrack B_{0e2}(\phi_2(t))sin\theta_2(t) \nonumber\\
 & \quad - 2\alpha B_2(\phi_2(t)) sin\theta_2 (t)cos\theta_2 (t) \nonumber\\
 & \quad - T_{dipole,\theta_2} - \alpha T_{dipole,\phi_2} \rbrack,
 \label{eq:theta_dynamics_bennett}
\end{align}
\begin{align}
\left(1+\alpha^2 \right) \cfrac{d\phi_2(t)}{dt} &= \cfrac{|\gamma|}{M_V} \cfrac{1}{sin\theta_2(t)} \lbrack \alpha B_{0e2}(\phi_2(t))sin\theta_2(t) \nonumber\\
	& \quad + 2 B_2(\phi_2(t)) sin\theta_2(t) cos\theta_2(t) \nonumber\\
	& \quad + \alpha T_{dipole,\theta_2} + T_{dipole,\phi_2}\rbrack \nonumber\\
  & \quad \hspace*{1in} (sin\theta \neq 0).
  \label{eq:phi_dynamics_bennett}
\end{align}

Note that in a very similar way the equations of dynamics for the other three nanomagnets can be derived.

On energy dissipation, we have one more component contributing to the total energy apart from the shape anisotropy and stress anisotropy energy, which is the energy due to dipole coupling. While calculating internal energy dissipation, the sum of the energy dissipations in all the four nanomagnets are considered but note that the dissipations in 1st and 4th nanomagnets are quite negligible since they don't quite switch and dissipation in 2nd nanomagnet is around twice that of in 3rd nanomagnet since 2nd nanomagnet switches a complete $180^\circ$, while the 3rd nanomagnet switches only about $90^\circ$. The instantaneous power dissipation for the 2nd nanomagnet can be determined as\cite{roy_phd}
\begin{equation}
P_{d,2}(t) = \cfrac{\alpha \, |\gamma|}{(1+\alpha^2) M_V} \, \left| \mathbf{T_{E,2}}(t) + \mathbf{T_{dipole,2}}(t)\right|^2.
\label{eq:power_dissipation_dipole}
\end{equation}
\noindent

\vspace*{5mm}
\begin{figurehere}
\centerline{\psfig{file=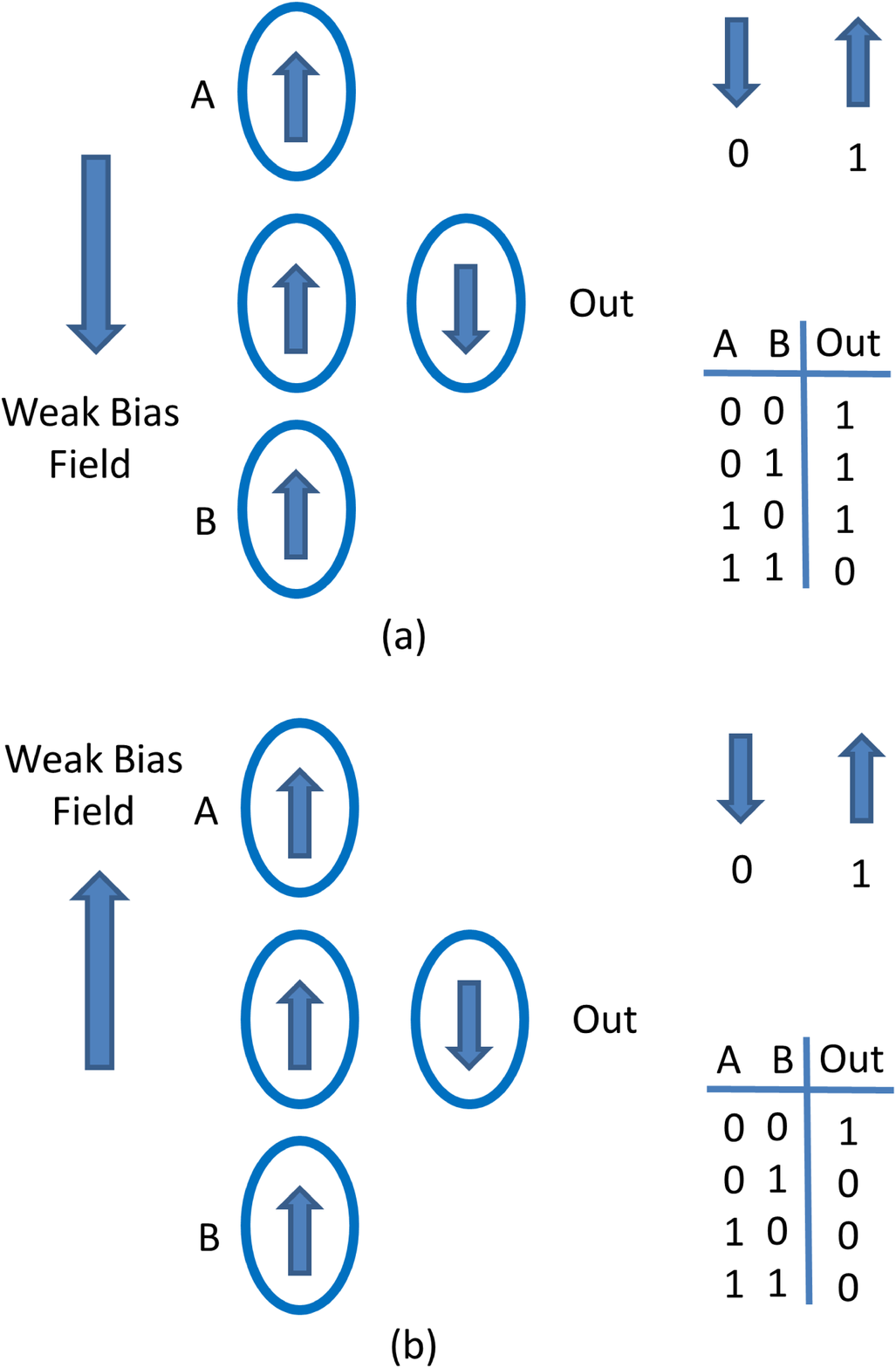,width=3.5in}}
\caption[Schematic of universal logic gates employing Magnetic Quantum Cellular Automata (MQCA) based architecture.]
{\label{fig:universal_logic_gates_dipole} Schematic of universal logic gates employing Magnetic Quantum Cellular Automata (MQCA) based architecture: (a) NAND gate, and (b) NOR gate. Note that a \emph{weak} bias field in the specified direction is required to break the tie when the input bits are different. The bias field must be weak enough so that it does not interfere in the operation when the input bits are $0$s for NAND gate and $1$s for NOR gate. (Reprinted with permission from Ref.~\phdref. Copyright 2012, Kuntal Roy.)}
\end{figurehere}

Note that we have not considered thermal flucutations and also have not applied any out-of-plane bias field so the term $\mathbf{T_M}(t)$ term as in Eq.~\eqref{eq:power_dissipation} is absent here. The power dissipations are integrated throughout the switching period to get the energy dissipation due to Gilbert damping. We have also considered `$CV^2$' energy dissipation, which can be significantly brought down by decreasing the stress since, stress is proportional to voltage applied, while sacrificing switching delay a bit. Since we have considered instantaneous ramp and stress is reversed during ramp-down phase, the `$CV^2$' energy dissipation is simply $3CV^2$ for the 2nd nanomagnet. 

We have considered Bennett clocking in an antiferromagnetically coupled horizontal wire for demonstration of magnetization dynamics in an array of multiferroic devices, however, a similar analysis is possible in the context of a ferromagnetically coupled vertical wire. We have not incorporated ramp rate effect or thermal fluctuations, which one needs to consider and analyze further. Universal logic gates (e.g. NAND and NOR gates) can also be constructed and analyzed using the very same model that includes dipole coupling. Fig.~\ref{fig:universal_logic_gates_dipole} depicts such possibilities. In general, magnetizations of an array of nanomagnets can be manipulated to implement computing in MQCA (Magnetic Quantum Cellular Automata) based architecture\cite{RefWorks:339}. Although such architecture has complexity of clocking each nanomagnet in the array, this is a \emph{regular} structure and circuits based on such structure can be designed systematically. Anyway, unconventional design of logic gates and building blocks for large-scale circuits using multiferroic composites can possibly be worked out too. Such designs may incur less complexity and possess better 	performance metrics than that of Bennett clocking mechanism.

\section{\label{sec:results}Simulation Results and Discussions}

In this Section, we review the simulation results for both single multiferroic devices\cite{roy11,roy11_6} and an array of multiferroic devices\cite{roy_phd,fasha11}. The performance metrics switching delay and energy dissipation are determined and trade-off between them is presented, i.e. if we want to make the switching faster, it would cost higher energy dissipation. Also, we determine the distributions of switching delay and energy dissipation, and number of successful switching events in the presence of room-temperature thermal fluctuations for a single multiferroic device.

\subsection{Single Multiferroic Device}

We consider the magnetostrictive layer to be made of polycrystalline Terfenol-D, nickel, or cobalt\cite{roy11}. Terfenol-D has 30 times higher magnetostriction coefficient in magnitude and it has the following material properties -- Young's modulus (Y): 8$\times$10$^{10}$ Pa, magnetostrictive coefficient ($(3/2)\lambda_s$): +90$\times$10$^{-5}$, saturation magnetization ($M_s$):  8$\times$10$^5$ A/m, and Gilbert's damping constant ($\alpha$): 0.1 (Refs.~\refcite{RefWorks:179,RefWorks:176,RefWorks:178,materials}). For the piezoelectric layer, we have considered lead-zirconate-titanate (PZT) having a dielectric constant of 1000. The maximum strain that can be generated in the PZT layer is 500 ppm\cite{RefWorks:170,RefWorks:563} and it would require a voltage of 66.7 mV because $d_{31}$=1.8$\times$10$^{-10}$ m/V for PZT\cite{pzt2}. The PZT layer is assumed to be four times thicker than the magnetostrictive layer so that any strain generated in it is transferred almost completely to the magnetostrictive layer\cite{RefWorks:154,roy11}. So the corresponding stress in Terfenol-D is the product of the generated strain ($500\times10^{-6}$) and the Young's modulus (8$\times$10$^{10}$ Pa). Hence, 40 MPa is the maximum stress that can be generated in the Terfenol-D nanomagnet. The strain-voltage relationship in PZT is actually {\it superlinear}  since $d_{31}$ increases with electric field\cite{RefWorks:563}. Hence, the voltage needed to produce 500 ppm strain in the Terfenol-D layer will be {\it less} than 66.7 mV and the energy dissipation would be a bit overestimated too.

We first review the results for low stress levels leading to slow switching speed (10-100 ns) and low energy dissipation\cite{roy11}. Fig.~\ref{fig:energy_harvesting_delay_energy_bothCV2andEtotal} (taken from Ref.~\refcite{roy11}) shows the energy dissipated in the switching circuit ($CV^2$) and the total energy dissipated ($E_{total}$) as functions of delay for three different materials (Terfenol-D, nickel, and cobalt) used as the magnetostrictive layer in the multiferroic nanomagnet. We solve magnetization dynamics to calculate the switching delay $\tau$ and also energy dissipation (`$CV^2$' dissipation and internal one $E_d$) for a given stress $\sigma$, and then we plot the switching delays and energy dissipations for different stress values. Terfenol-D incurs much less energy dissipation than the other two materials because it has much higher magnetostriction coefficient requiring a less stress level to generate a certain stress anisotropy. For Terfenol-D, the stress required to switch in 100 ns is 1.92 MPa and that required to switch in 10 ns is 2.7 MPa. Note that for a stress of 1.92 MPa, the stress anisotropy energy $B_{stress}$ is 32.7 $kT$  while for 2.7 MPa, it is 46.2 $kT$. Since the stress level is assumed to be applied instantaneously, as expected, the energy dissipation numbers are larger than the shape anisotropy barrier of $\sim$32 $kT$.  A larger excess energy is needed to switch faster signifying the delay-energy trade-off. The energy dissipated and lost as heat in the switching circuit ($CV^2$) is only 12 $kT$ for a delay of 100 ns, while that is 23.7 $kT$ for a delay of 10 ns. The total energy dissipated is 45 $kT$ for switching delay of 100 ns and 70 $kT$ for switching delay of 10 ns. 

\vspace*{5mm}
\begin{figurehere}
\centerline{\psfig{file=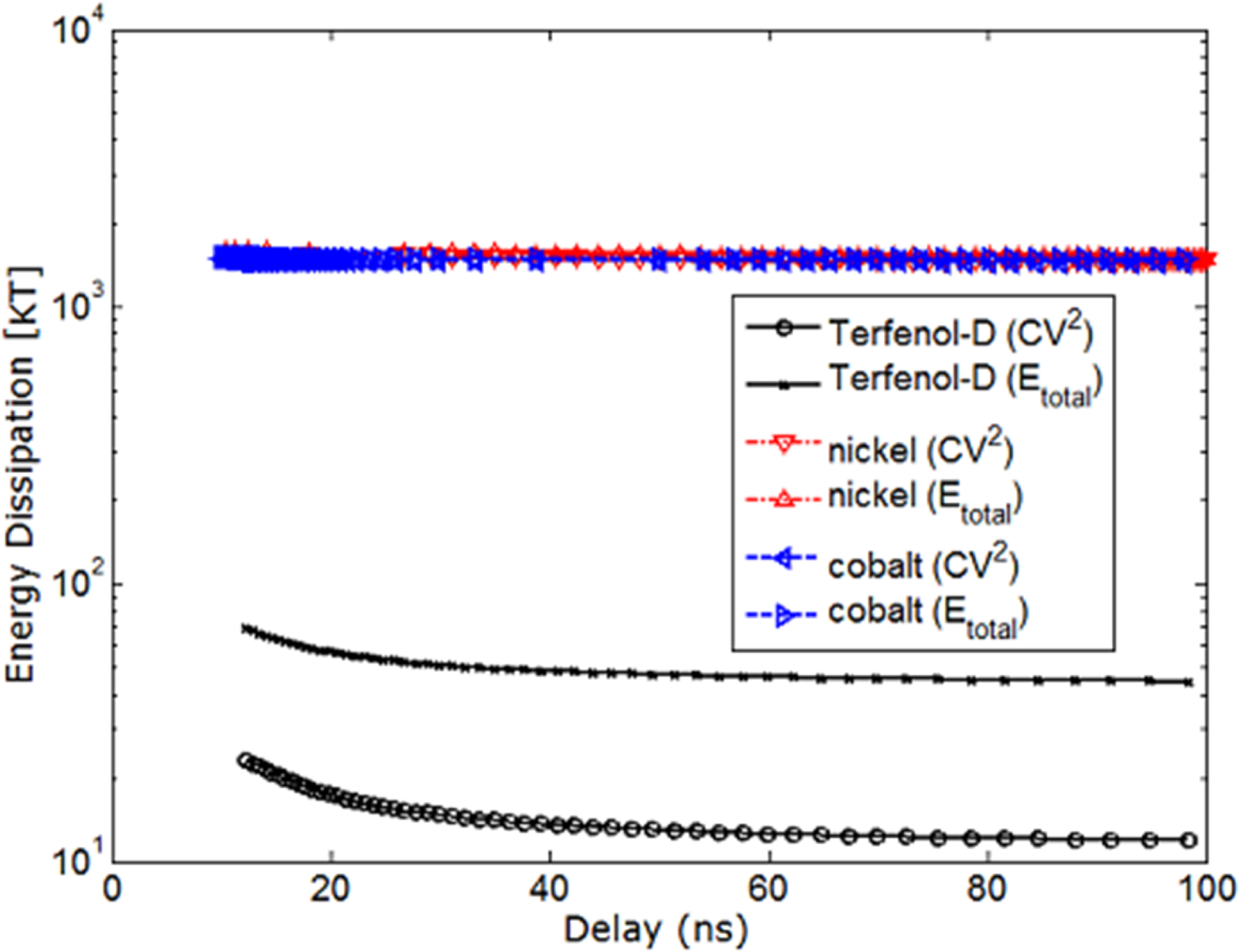,width=3.5in}}
\caption{\label{fig:energy_harvesting_delay_energy_bothCV2andEtotal} Energy dissipated in the switching circuit ($CV^2$) and the total energy dissipated ($E_{total}$) as functions of delay for three different materials used as the magnetostrictive layer in the multiferroic nanomagnet. (Reprinted with permission from Ref.~\aplref. Copyright 2011, AIP Publishing LLC.)}
\end{figurehere}

With a nanomagnet density of 10$^{10}$ cm$^{-2}$ in a memory or logic chip, and if we consider 10\% of the nanomagnets switch at any given time (10\% activity level), the dissipated power density would have been only 2 mW/cm$^2$ to switch in 100 ns and 30 mW/cm$^2$ to switch in 10 ns. Such {\it extremely low power} and yet {\it high density} magnetic logic and memory systems can be powered by existing energy harvesting systems\cite{roundy03,anton,lu,jeon} that harvest energy from the environment without the need for an external battery. These processors are deemed to be suitable for implantable medical devices, e.g. those implanted in a patient's brain that monitor brain signals to warn of impending epileptic seizures. They can run on energy harvested from the patient's body motion. For such applications, 10-100 ns switching delay is adequate.  

We now review multiferroic devices for higher stress levels and fast sub-nanosecond switching speed, which is particularly important for logic and computing purposes\cite{roy11_6}. We performed simulations in the presence of room-temperature thermal fluctuations and we consider only Terfenol-D as magnetostrictive material since it has much higher magnetostrictive coefficient (than nickel and cobalt) thus being fruitful in resisting the adverse effects of thermal fluctuations.

Fig.~\ref{fig:thermal_theta_phi_distribution_terfenolD_mag_field} plots the distributions of initial angles $\theta_{initial}$ and $\phi_{initial}$ in the presence of thermal fluctuations and a bias magnetic field applied along the out-of-plane direction (+$x$-axis). The bias field has shifted the peak of $\theta_{initial}$ exactly from the easy axis ($\theta = 180^{\circ}$) as shown in Fig.~\ref{fig:thermal_theta_phi_distribution_terfenolD_mag_field}(a). The $\phi_{initial}$ distribution (see Fig.~\ref{fig:thermal_theta_phi_distribution_terfenolD_mag_field}(b)) has two peaks and resides mostly within the interval [-90$^\circ$,+90$^\circ$] since the bias magnetic field is applied in the +$x$-direction. Because the magnetization vector starts out from near the south pole ($\theta \simeq 180^\circ$) when stress is turned on, the effective torque on the magnetization [$\sim\mathbf{M} \times \mathbf{H}$, where $\mathbf{M}$ is the magnetization and $\mathbf{H}$ is the effective field] due to the +$x$-directed magnetic field is such that the magnetization prefers the $\phi$-quadrant (0$^\circ$,90$^\circ$) over the $\phi$-quadrant (270$^\circ$,360$^\circ$), which is the reason for the asymmetry in the two distributions of $\phi_{initial}$. Consequently, when the magnetization vector starts out from $\theta \simeq 180^{\circ}$, the initial azimuthal angle $\phi_{initial}$ is more likely to be in the quadrant (0$^\circ$,90$^\circ$) than in the quadrant (270$^\circ$,360$^\circ$).

We assume that when a \emph{compressive} stress is applied to initiate switching (since Terfenol-D has \emph{positive} magnetostrictive coefficient), the magnetization vector starts out from near the south pole ($\theta \simeq 180^\circ$) with a certain ($\theta_{initial}$,$\phi_{initial}$) picked from the initial angle distributions at room-temperature. Stress is ramped up linearly and kept constant until the magnetization reaches $x$-$y$ plane ($\theta = 90^\circ$). Then stress is ramped down at the same rate at which it was ramped up, and reversed in magnitude to aid switching. 
The magnetization dynamics ensures that $\theta$ continues to rotate towards $0^{\circ}$. When $\theta$ becomes $ \leq 5^\circ$, switching is deemed to have completed. A moderately large number (10,000) of simulations, with their corresponding ($\theta_{initial}$,$\phi_{initial}$) picked from the initial angle distributions, are performed for each value of stress and ramp duration to generate the simulation results in the presence of thermal fluctuations. 

Fig.~\ref{fig:thermal_stress_voltage_success_ramp_time_mag} shows the switching probability as a function of stress levels (10-30 MPa) and as well as voltage applied across the piezoelectric layer for different ramp durations (60 ps, 90 ps, 120 ps)\cite{roy11_2,RefWorks:430} at room temperature (300 K). The minimum stress needed to switch the magnetization with $\sim$100\% probability at 300 K is $\sim$14 MPa for 60 ps ramp duration and $\sim$17 MPa for 90 ps ramp duration. For higher ramp duration of 120 ps, the curve is non-monotonic, which we will discuss later.

\vspace*{5mm}
\begin{figurehere}
\centerline{\psfig{file=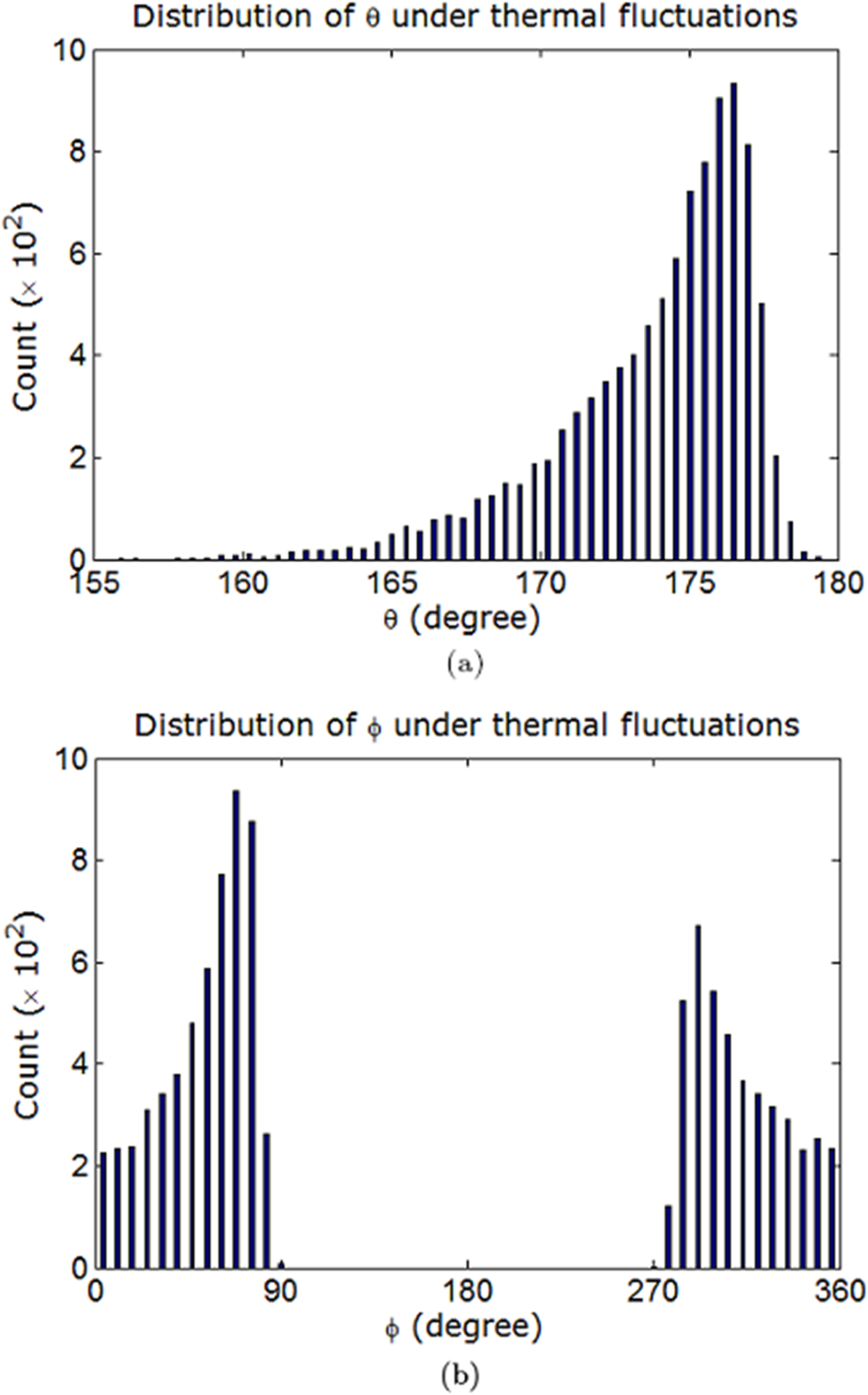,width=3.5in}}
\caption{\label{fig:thermal_theta_phi_distribution_terfenolD_mag_field} Distribution of polar angle $\theta_{initial}$ and azimuthal angle $\phi_{initial}$ due to thermal fluctuations at room temperature (300 K) when a magnetic field of flux density 40 mT is applied along the out-of-plane hard axis (+$x$-direction).
(a) Distribution of polar angle $\theta_{initial}$ at room temperature (300 K). The mean of the distribution is $173.7^\circ$, and the most likely value is 175.8$^{\circ}$.
(b) Distribution of the azimuthal angle $\phi_{initial}$ due to thermal fluctuations at room temperature (300 K). There are two distributions with  peaks centered at $\sim$65$^\circ$ and $\sim$295$^\circ$. (Reprinted with permission from Ref.~\japref. Copyright 2012, AIP Publishing LLC.)}
\end{figurehere}

At low stress levels (10-20 MPa), the switching probability increases with stress, regardless of the ramp rate. This happens because a higher stress can more effectively counter the adverse effects of thermal fluctuations to facilitate switching, and hence increases the success rate of switching. This feature is independent of the ramp rate for lower stress levels. However, for higher stress levels accompanied by a higher ramp duration, the switching dynamics is complex, which we will explain onwards.

\vspace*{5mm}
\begin{figurehere}
\centerline{\psfig{file=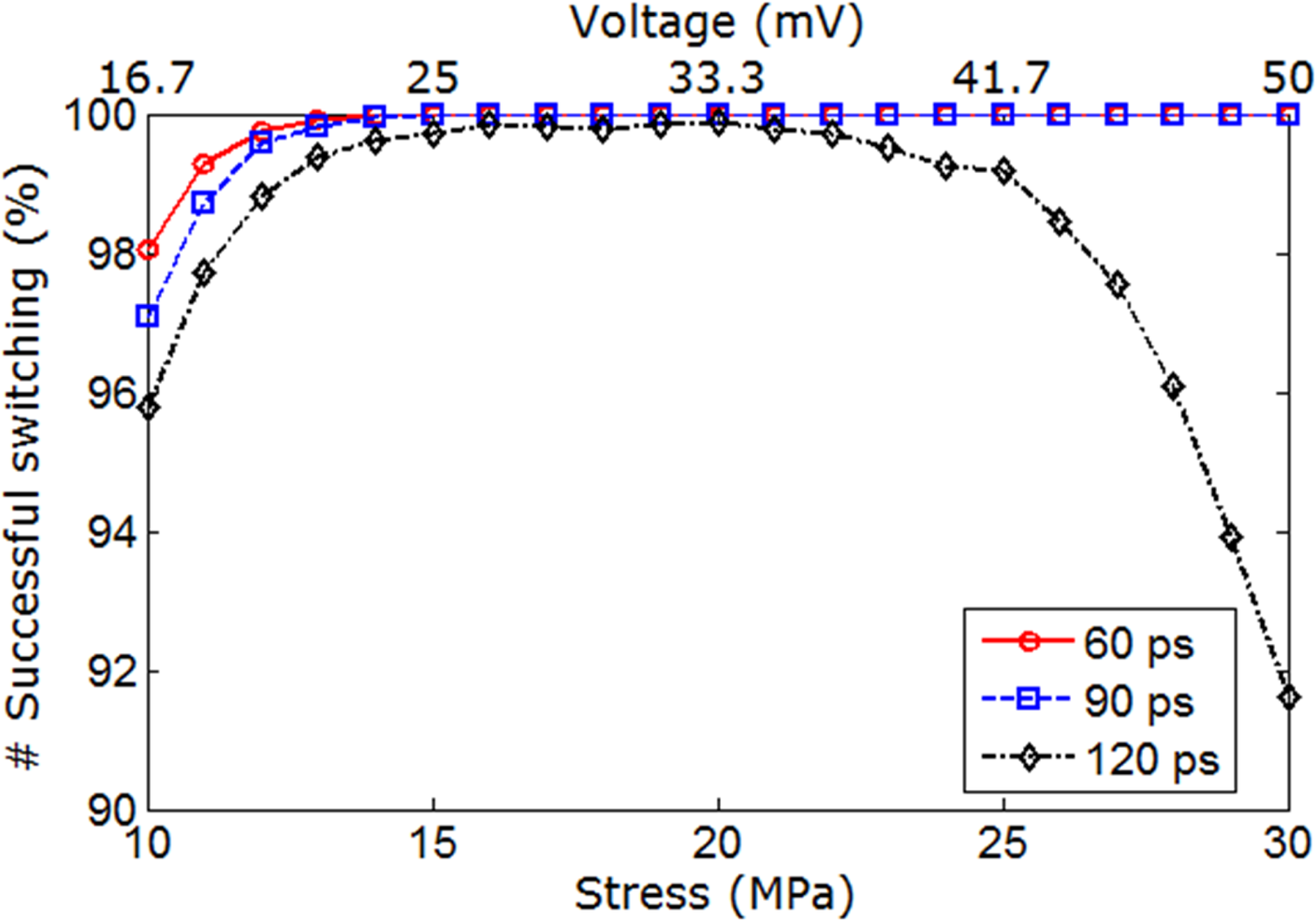,width=3.5in}}
\caption{\label{fig:thermal_stress_voltage_success_ramp_time_mag} Percentage of
successful switching events among the simulated switching trajectories (or the switching probability) at room temperature in a Terfenol-D/PZT multiferroic nanomagnet versus (lower axis) stress (10-30 MPa) and (upper axis) voltage applied across the piezoelectric layer, for different ramp durations (60 ps, 90 ps, 120 ps). The stress at which switching becomes $\sim$100\% successful increases with ramp duration. For large ramp duration (120 ps) or slow ramp rate, $\sim$100\% switching probability is unachievable. (Reprinted with permission from Ref.~\japref. Copyright 2012, AIP Publishing LLC.)}
\end{figurehere}

We will now describe the significance of ramp-rate on success probability. When stress (\emph{compressive} stress for Terfenol-D) is made active, magnetization traverses from $\theta \simeq 180^{\circ}$ towards $x$-$y$ plane ($\theta = 90^{\circ}$). Once the magnetization vector crosses the $x$-$y$ plane, the stress needs to be withdrawn as soon as possible. This is because the stress forces the energy minimum to remain at $\theta = 90^{\circ}$, which will make the magnetization linger around $\theta = 90^{\circ}$ instead of rotating towards the desired direction $\theta \simeq 0^{\circ}$. This is why stress must be removed or reversed immediately upon crossing the $x$-$y$ plane so that the energy minimum moves to $\theta = 0^{\circ}, 180^{\circ}$. Then the magnetization vector rotates towards $\theta = 0^{\circ}$ rather than $\theta = 180^{\circ}$ due to inherent switching dynamics\cite{roy11_5}. If the removal rate is fast, then the success probability is high since the harmful stress does not stay active long enough to cause significant backtracking of the magnetization vector towards $\theta = 90^{\circ}$. However, if the ramp rate is too slow, then significant backtracking can possibly occur whereupon the magnetization vector may return to the $x$-$y$ plane. Then thermal torque decides the fate whether magnetization backtracks towards $\theta \simeq 180^{\circ}$, causing switching failure or switches successfully towards $\theta \simeq 0^{\circ}$; in general there is 50\% switching probability then. Hence the switching probability drops with decreasing ramp rate.

\vspace*{5mm}
\begin{figurehere}
\centerline{\psfig{file=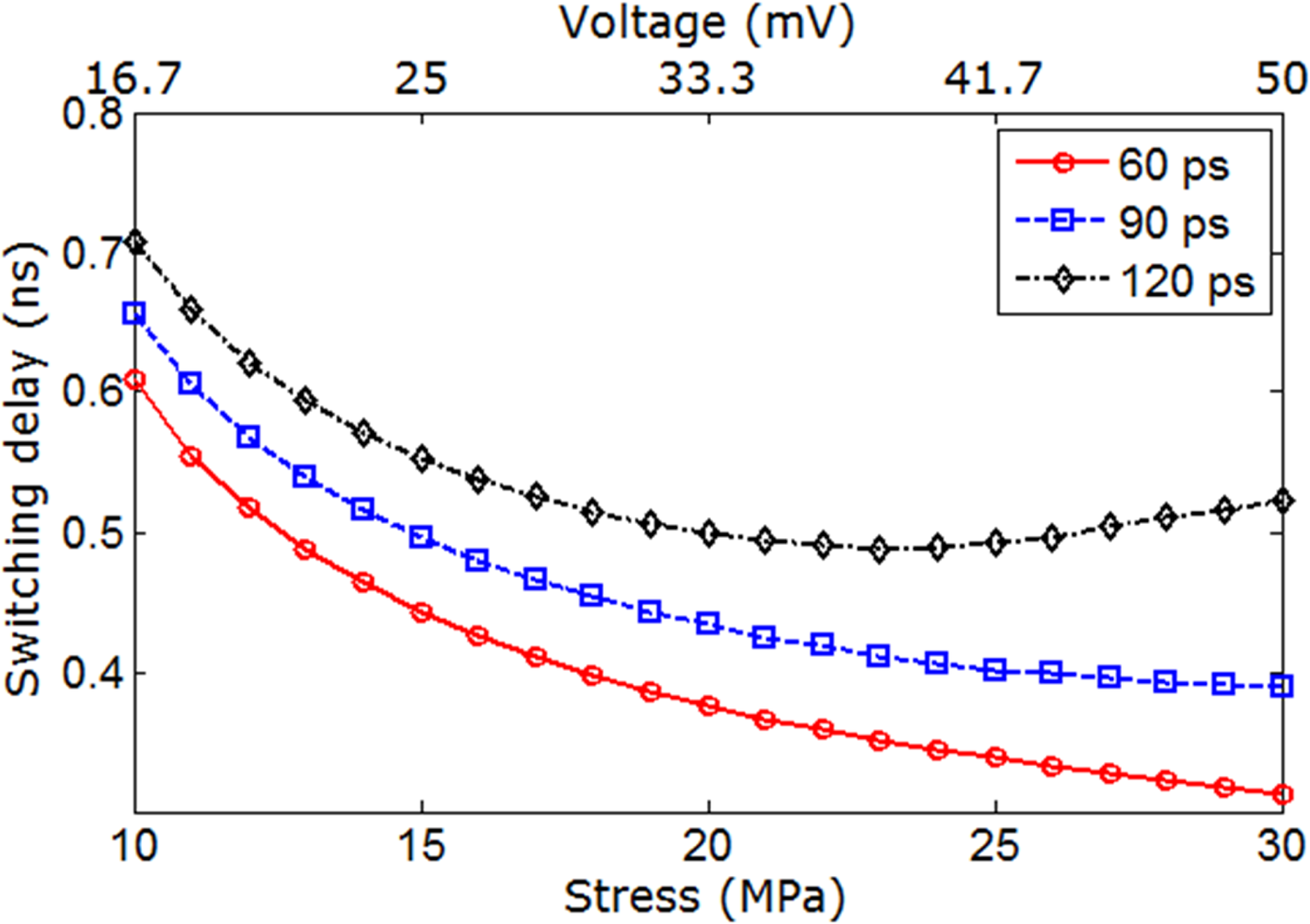,width=3.5in}}
\caption{\label{fig:thermal_mean_stress_voltage_delay_mag} The thermal mean of the switching delay (at 300 K) versus (lower axis) stress (10-30 MPa) and (upper axis) voltage applied across the piezoelectric layer, for different ramp durations (60 ps, 90 ps, 120 ps). Switching may fail at low stress levels and also at high stress levels for long ramp durations. Failed attempts are excluded when computing the mean. (Reprinted with permission from Ref.~\japref. Copyright 2012, AIP Publishing LLC.)} 
\end{figurehere}

A similar explanation is also applicable for the non-monotonic stress dependence of the switching probability when the ramp rate is slow (ramp duration of 120 ps). When $\theta$ is in the quadrant [180$^{\circ}$, 90$^{\circ}$], a higher stress is helpful since it provides a larger torque to move towards the $x$-$y$ plane, but when $\theta$ is in the quadrant [90$^{\circ}$, 0$^{\circ}$], a higher stress is harmful since it increases the chance of backtracking, particularly when the ramp-down rate is slow. These two counteracting effects are the reason for the non-monotonic dependence of the success probability on stress in the case of the slowest ramp rate. At higher stress levels accompanied by a slow ramp rate, it causes a significant amount of backtracking causing the switching probability to drop fast. For slow ramp rate (ramp duration of 120 ps), we have not observed 100\% switching probability at any stress for the 10,000 simulations performed.

Fig.~\ref{fig:thermal_mean_stress_voltage_delay_mag} plots the thermal mean (averaged from 10,000 simulations) switching delay versus stress for different ramp durations. Specifically, only successful switching events are considered here since we do not have a value of switching delay for an unsuccessful switching event. For ramp durations of 60 ps and 90 ps, the switching delay decreases with increasing stress since the torque, which rotates the magnetization, increases when stress increases. However, for 120 ps ramp duration, the dependence is non-monotonic, due to same reason causing the non-monotonicity in Fig.~\ref{fig:thermal_stress_voltage_success_ramp_time_mag}. For a certain stress, decreasing the ramp duration (or increasing the ramp rate) decreases the switching delay because the stress reaches its maximum value quicker and hence switches the magnetization faster.

\vspace*{5mm}
\begin{figurehere}
\centerline{\psfig{file=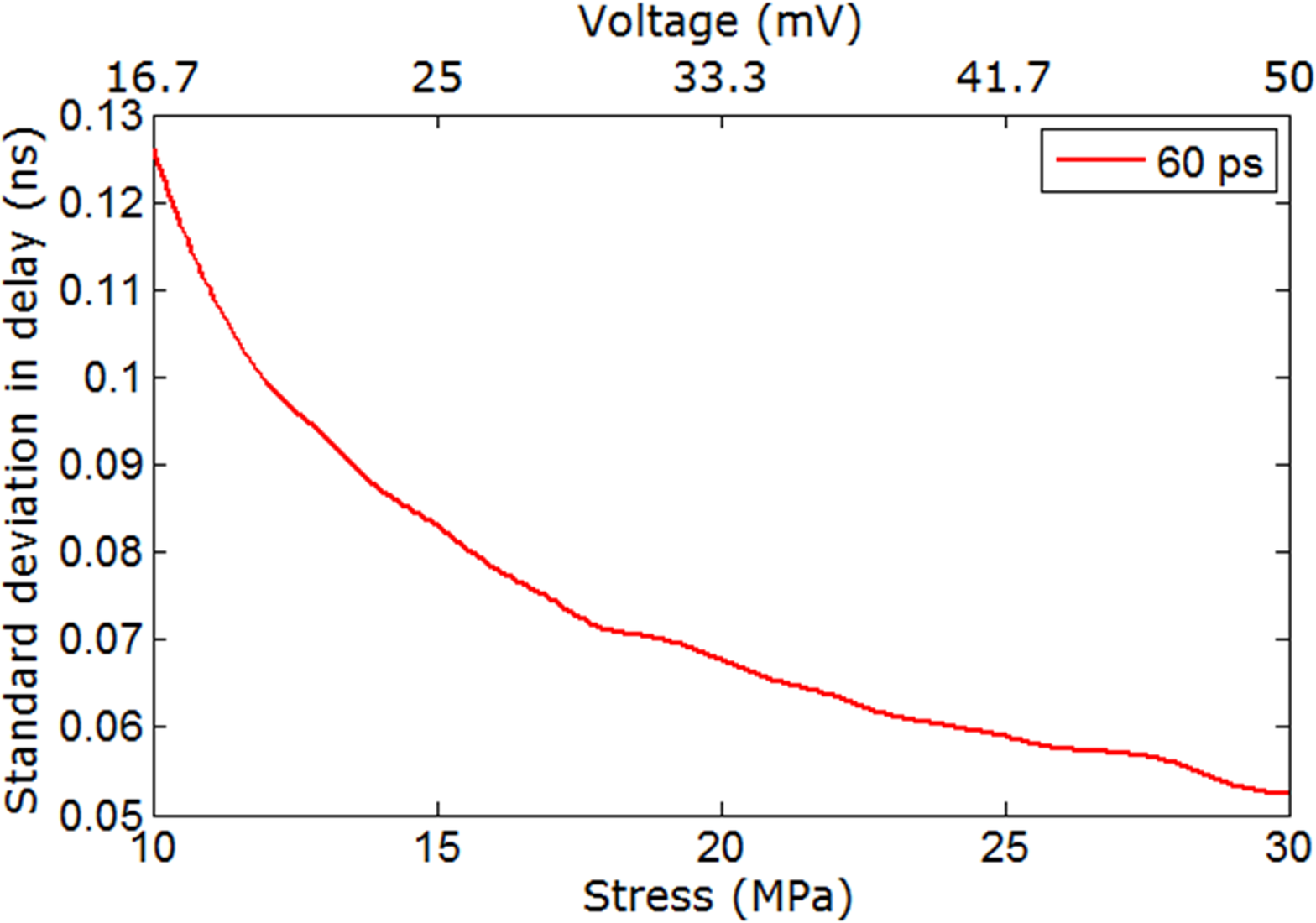,width=3.5in}}
\caption{\label{fig:thermal_stress_voltage_delay_std} The standard deviations in switching delay versus (lower axis) stress (10-30 MPa) and (upper axis) voltage applied across the piezoelectric layer for 60 ps ramp duration at 300 K. We consider only the successful switching events in determining the standard deviations. The standard deviations in switching delay  for other ramp durations are of similar magnitudes and show similar trends. (Reprinted with permission from Ref.~\japref. Copyright 2012, AIP Publishing LLC.)}
\end{figurehere}

Fig.~\ref{fig:thermal_stress_voltage_delay_std} plots the standard deviation in switching delay versus stress for 60 ps ramp duration. The results for other ramp durations are similar and hence are not shown for brevity. At higher values of stress, the torque due to stress is stronger and dominates over the random thermal torque, which causes the spread in the switching delay. This decreases the standard deviation in switching delay with increasing stress and thus the switching delay distribution gets more peaked as we increase the stress.

Fig.~\ref{fig:thermal_mean_stress_voltage_energy_mag} plots the thermal mean of the total energy dissipated to switch the magnetization as a function of stress and voltage across the piezoelectric layer for different ramp durations. To understand these curves, we need to consider first the trend of average power dissipation ($E_{total}/\tau$), which increases with stress for a given ramp duration and decreases with increasing ramp duration for a given stress. More stress corresponds to more `$CV^2$' dissipation and also more internal power dissipation because it results in a higher torque. Slower switching decreases the power dissipation since it makes the switching more adiabatic. Note that the switching delay curves show the opposite trend (see Fig.~\ref{fig:thermal_mean_stress_voltage_delay_mag}) than that of average power dissipation. At a lower ramp rate (higher ramp duration), the average power dissipation $E_{total}/\tau$ is always smaller than that of a higher ramp rate, but the switching delay does not decrease as fast as with higher values of stress (in fact switching delay may increase for higher ramp duration), which is why the energy dissipation curves in  Fig.~\ref{fig:thermal_mean_stress_voltage_energy_mag} exhibit the cross-overs. 

\vspace*{5mm}
\begin{figurehere}
\centerline{\psfig{file=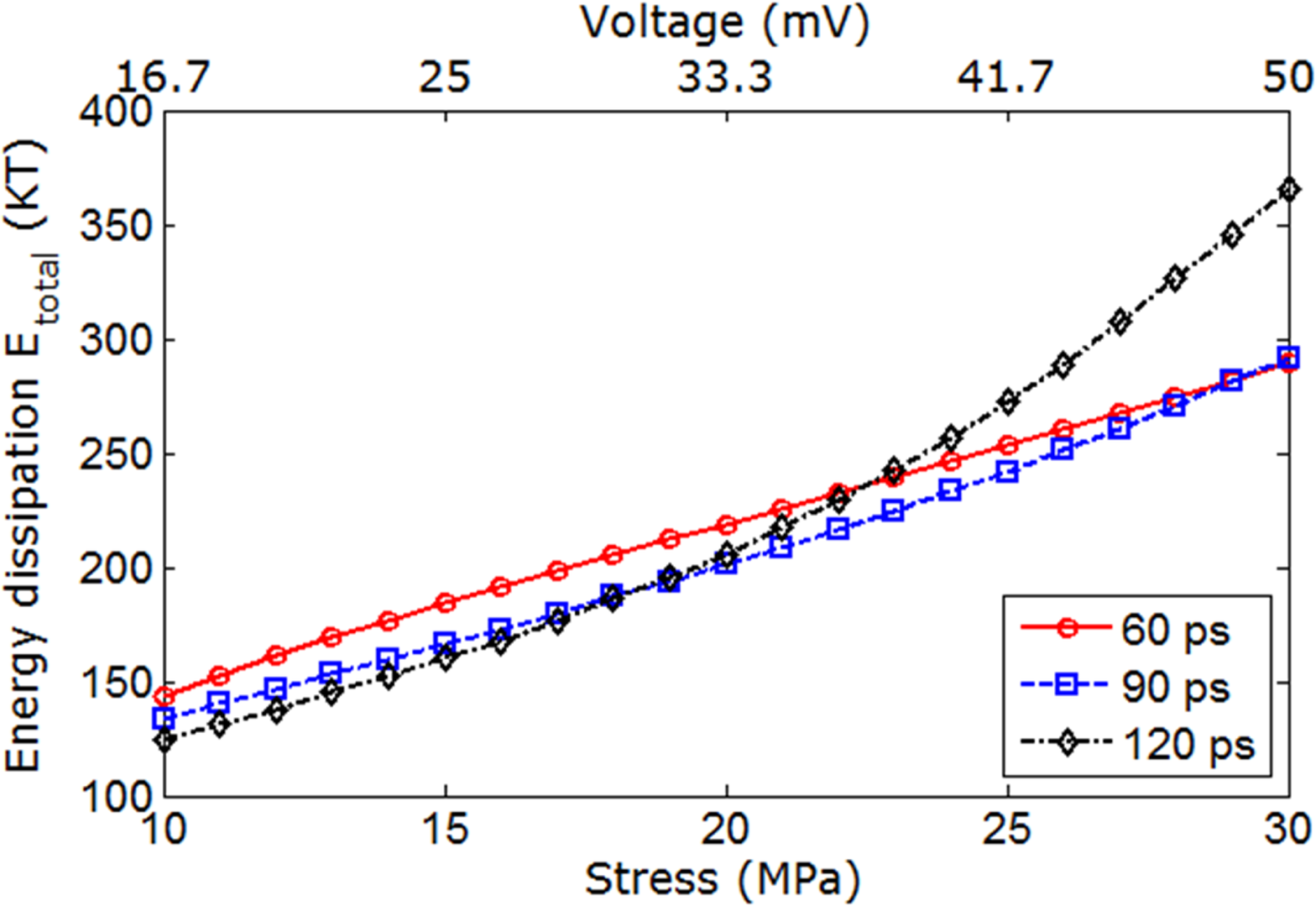,width=3.5in}}
\caption{\label{fig:thermal_mean_stress_voltage_energy_mag} Thermal mean of the total energy dissipation versus (lower axis) stress (10-30 MPa) and (upper axis) voltage across the piezoelectric layer for different ramp durations (60 ps, 90 ps, 120 ps). Once again, failed switching attempts are excluded when computing the mean. (Reprinted with permission from Ref.~\japref. Copyright 2012, AIP Publishing LLC.)}
\end{figurehere}

Fig.~\ref{fig:thermal_stress_voltage_energy_CV2} plots the `$CV^2$' energy dissipation in the switching circuitry versus stress and the voltage applied across the PZT layer. Increasing stress requires increasing the voltage $V$, which is why the `$CV^2$' energy dissipation increases rapidly with stress. This dissipation however is a small fraction of the total energy dissipation ($<$ 15\%) since a very small voltage is required to switch the magnetization of a multiferroic nanomagnet with stress. The `$CV^2$' dissipation decreases when the ramp duration increases because then the switching becomes more `adiabatic' and hence less dissipative. This component of the energy dissipation would have been several orders of magnitude higher had we switched the magnetization with an external magnetic field\cite{RefWorks:142,RefWorks:124} or spin-transfer torque induced by spin-polarized current\cite{RefWorks:7}. This makes the stress-induced switching of magnetization promising for pursuing our future nanoelectronics. 

\vspace*{5mm}
\begin{figurehere}
\centerline{\psfig{file=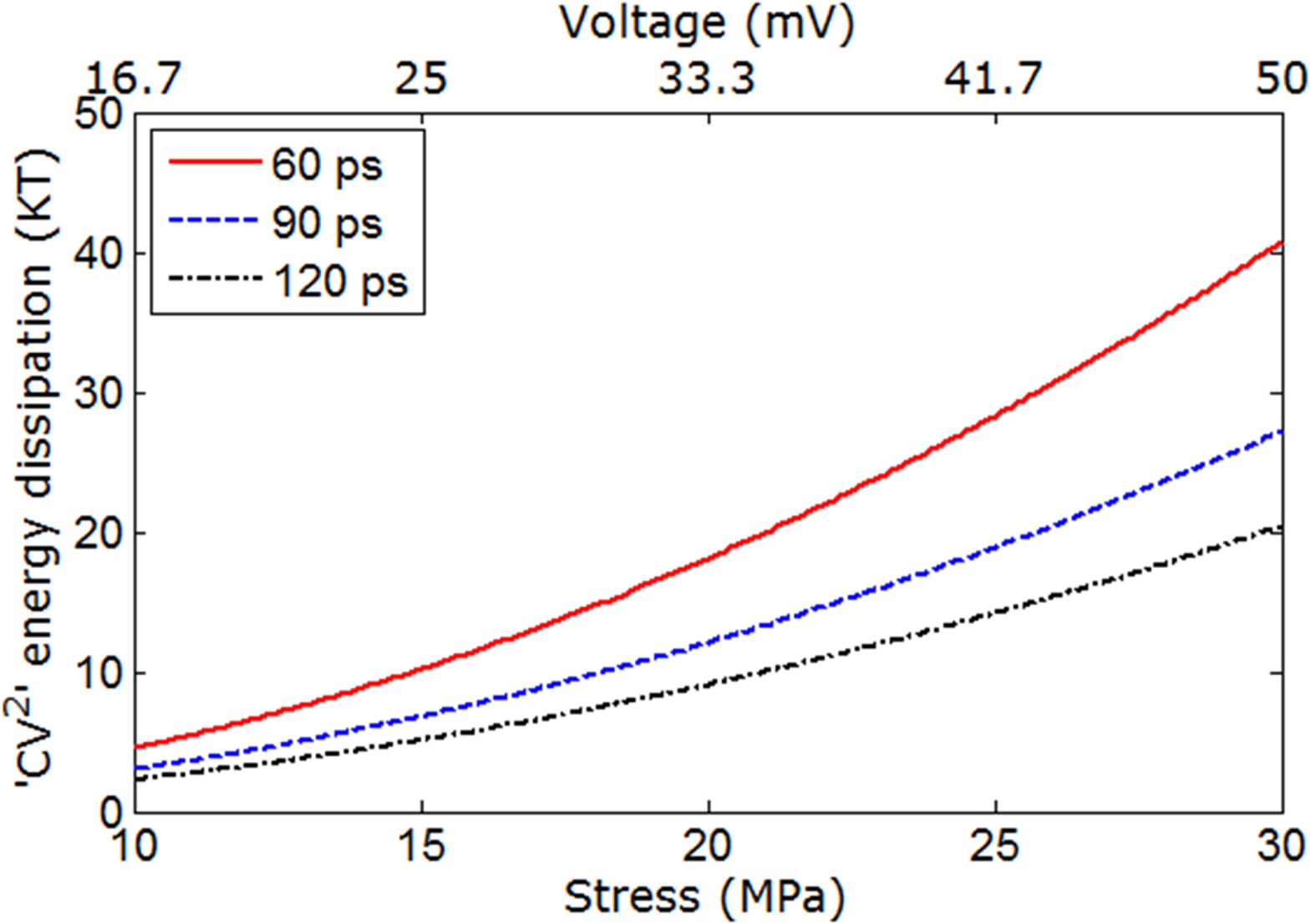,width=3.5in}}
\caption{\label{fig:thermal_stress_voltage_energy_CV2} The `$CV^2$' energy dissipation in the external circuit as a function of (lower axis) stress and (upper axis) voltage applied across the PZT layer for different ramp durations. The dependence on voltage is not exactly quadratic since the voltage is not applied abruptly, but instead ramped up gradually and linearly in time. (Reprinted with permission from Ref.~\japref. Copyright 2012, AIP Publishing LLC.)}
\end{figurehere}

Fig.~\ref{fig:thermal_distribution_delay_energy_60ps_15MPa_mag} plots the switching delay and energy distributions in the presence of room-temperature thermal fluctuations for 15 MPa stress and 60 ps ramp duration. The high-delay tail in Fig.~\ref{fig:thermal_distribution_delay_energy_60ps_15MPa_mag}(a) is in general associated with those switching trajectories that start close to $\theta = 180^\circ$. In such trajectories, the starting torque is vanishingly small as explained earlier, which makes the switching sluggish at the beginning. During this time, switching also becomes susceptible to backtracking because of thermal fluctuations, which may also increase the delay further. However, it may well happen that random thermal torque quite facilitates a switching trajectory even it gets started very close to the easy axis making the \emph{net} switching fast. Note that there was not a single event where the delay exceeded 1 ns out of 10,000 simulations of switching trajectories, showing that the probability of that happening is less than 0.01\% and probably far less than 0.01\%. We have analyzed the cause of such high-delay tail and have been able to reduce it by application of an out-of-plane bias field. The magnitude of bias field can be calibrated to reduce the extent of such tail further. Since the energy dissipation is the product of the power dissipation and the switching delay, similar behavior is found in Fig.~\ref{fig:thermal_distribution_delay_energy_60ps_15MPa_mag}(b).

\vspace*{5mm}
\begin{figurehere}
\centerline{\psfig{file=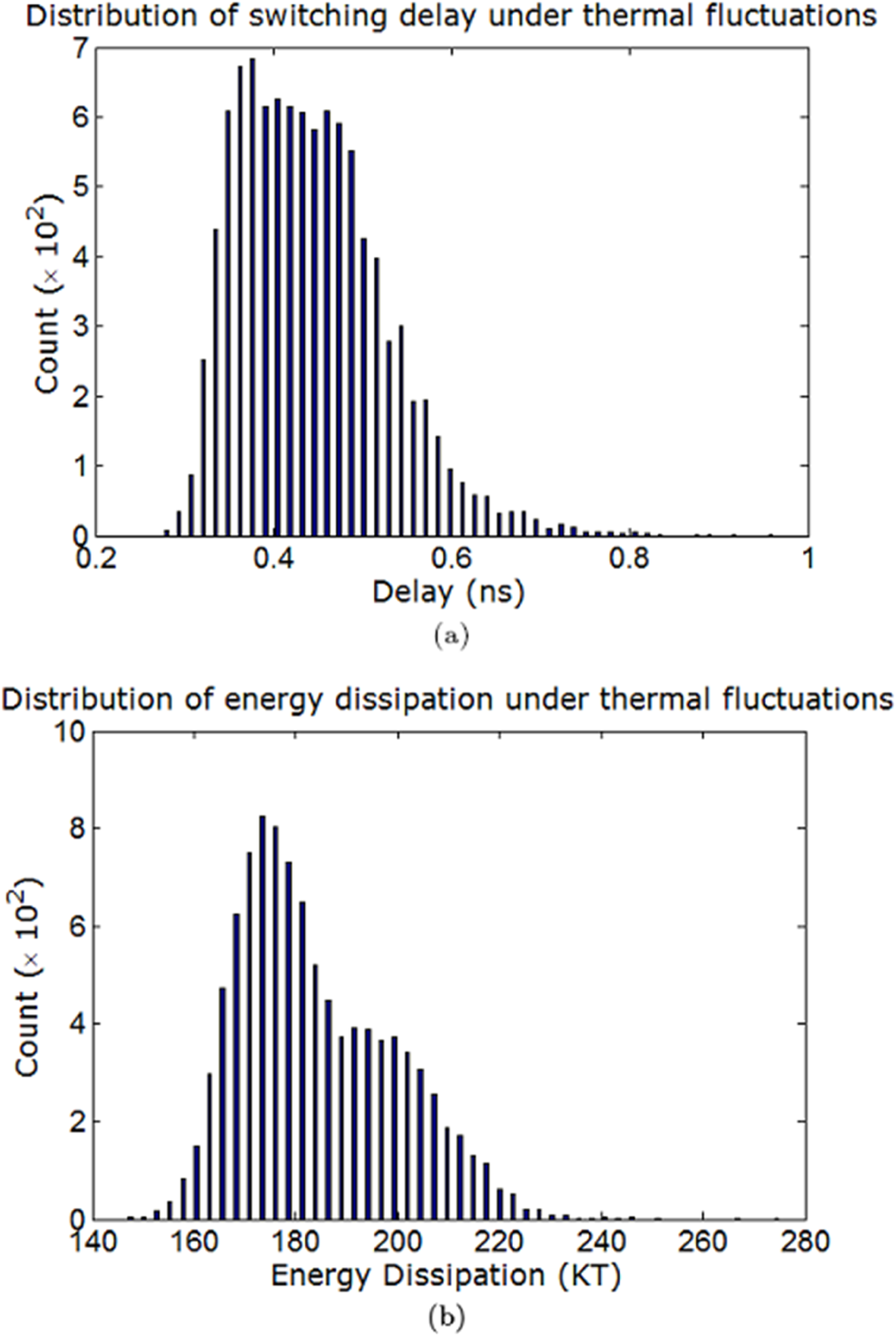,width=3.5in}}
\caption{\label{fig:thermal_distribution_delay_energy_60ps_15MPa_mag} Delay and energy distributions for 15 MPa applied stress and 60 ps ramp duration at room temperature (300 K). (a) Distribution of the switching delay. The mean and standard deviation of the distribution are 0.44 ns and 83 ps, respectively. (b) Distribution of energy dissipation. The mean and standard deviation of the distribution are 184 $kT$ and 15.5 $kT$ at room temperature, respectively. (Reprinted with permission from Ref.~\japref. Copyright 2012, AIP Publishing LLC.)}
\end{figurehere}

\subsection{Array of Multiferroic Devices}

Here we review the simulation results of unidirectional information propagation through an array of multiferroic devices using Bennett clocking mechanism\cite{roy_phd,fasha11}. We do not consider thermal fluctuations, nonetheless in all the simulations, the initial orientation of the magnetization vector is assumed as $\theta=175^\circ$ and $\phi=90^\circ$. Stress is applied instantaneously and we solve Eqs.~\eqref{eq:theta_dynamics_bennett} and~\eqref{eq:phi_dynamics_bennett} for the 2nd nanomagnet (and similar equations for the other three nanomagnets) at each time step. Once  $\theta$ becomes 90$^\circ$, stress is \emph{reversed} instantaneously and we follow the magnetization vector in time until $\theta$ becomes $\leq5^{\circ}$. At that point, switching is deemed to have completed.

Fig.~\ref{fig:Bennett_terfenolD_5d2MPa} shows the magnetization dynamics for all the nanomagnets when propagating a signal unidirectionally in a chain of four Terfenol-D/PZT multiferroic nanomagnets using Bennett clocking mechanism for the case of 5.2 MPa stress. When 5.2 MPa compressive stress is applied on the 2nd and 3rd nanomagnets and after their magnetizations come to their hard axes, stress is reversed on the 2nd nanomagnet to relax its magnetization towards its desired state. In the Fig.~\ref{fig:Bennett_terfenolD_5d2MPa}(b), note that the applied stress has deflected the azimuthal angle of magnetization $\phi$ of the 2nd nanomagnet in the quadrant ($90^\circ,180^\circ$) while that for the 3rd nanomagnet is deflected in the quadrant ($0^\circ,90^\circ$). These out-of-plane excursions aid the switching to be very fast in sub-nanosecond; the same physics as for a single multiferroic device\cite{roy11,roy11_6} applies here too. Had we not considered such out-of-plane excursion, the switching delay would have been a couple of magnitudes larger, which signifies that steady-state analysis without considering out-of-plane motion\cite{RefWorks:154} is neither qualitatively nor quantitatively accurate\cite{roy_phd,fasha11}. 

The stress of 5.2 MPa has only shifted the magnetizations out-of-plane by $5^\circ$ during $90^\circ$ switching, i.e. when the magnetizations come near to their hard axes. Upon reversing the stress on the 2nd nanomagnet, its magnetization rotates out-of-plane more, but this time at the very end, it has come back to its plane because of the dipole coupling with the 3rd nanomagnet, which tries to align the 2nd nanomagnet's magnetization with its own magnetization. Finally, note that the magnetizations of 1st and 4th nanomagnets remain quite unchanged because no stress is applied on these two nanomagnets, the slight changes in the directions of the magnetizations occurred because of dipole coupling effect with the neighboring nanomagnets.

Fig.~\ref{fig:Bennett_terfenolD_30MPa} shows magnetization dynamics for Bennett clocking with 30 MPa stress, rather than 5.2 MPa stress as we have presented earlier. With high stress, magnetizations have deflected out-of-plane more ($\sim$$10^\circ$) than that for the lower stress while reaching $\theta = 90^\circ$ and the magnetization of 2nd nanomagnet has executed a precessional motion before completing switching. Apparently, almost half of the time taken during switching, magnetization experiences such unfruitful motion. To reduce such unfruitful motion, one can just use a lower stress level. If we apply 10 MPa stress, rather than the high stress 30 MPa, and magnetization does not execute any precessional motion and the switching delay increases just a bit from 0.32 ns to 0.34 ns, while energy dissipation decreases 10 times due to a lower stress level. Thus it is possible to engineer the stress levels to achieve good performance metrics for application purposes.

\vspace*{5mm}
\begin{figurehere}
\centerline{\psfig{file=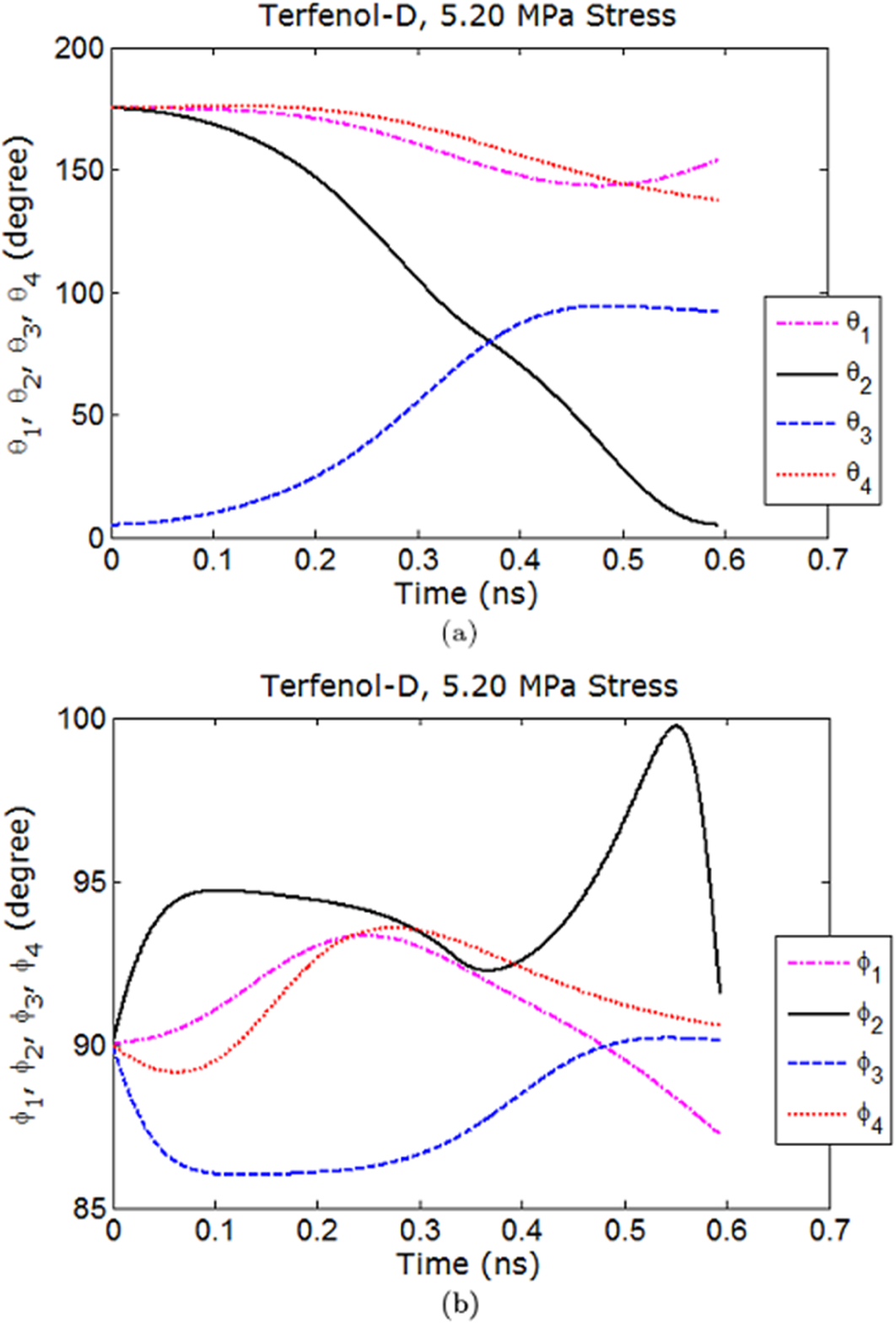,width=3.5in}}
\caption{\label{fig:Bennett_terfenolD_5d2MPa} Magnetization dynamics for Bennett clocking in a chain of four Terfenol-D/PZT multiferroic nanomagnets with stress 5.2 MPa and assuming instantaneous ramp: (a) polar angle $\theta$ versus time, and (b) azimuthal angle $\phi$ over time while switching occurs, i.e. during the time $\theta_2$ changes from 175$^{\circ}$ to 5$^{\circ}$. (Reprinted with permission from Ref.~\phdref. Copyright 2012, Kuntal Roy.)}
\end{figurehere}

Similarly, we can analyze and simulate logic gates (see Fig.~\ref{fig:universal_logic_gates_dipole}) and bigger circuits using multiferroic composites. Since the 3-phase clocking circuitry for Bennett clocking may be complex for application purposes, any unconventional design of logic gates and building blocks can be worked out and simulated too, however, based on the same technique presented here. 

\vspace*{5mm}
\begin{figurehere}
\centerline{\psfig{file=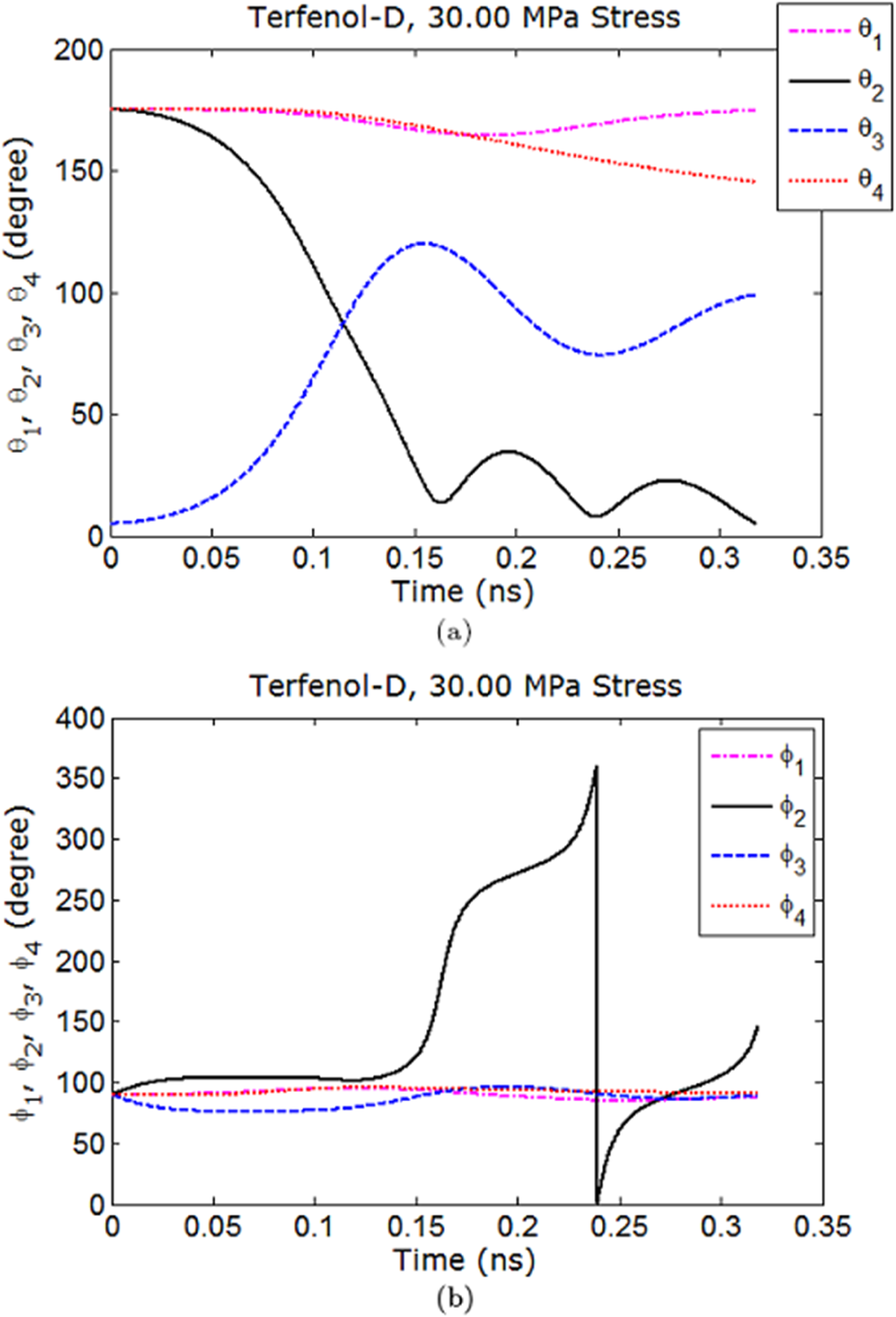,width=3.5in}}
\caption{\label{fig:Bennett_terfenolD_30MPa} Magnetization dynamics for Bennett clocking in a chain of four Terfenol-D/PZT multiferroic nanomagnets with stress 30 MPa and assuming instantaneous ramp: (a) polar angle $\theta$ versus time, and (b) azimuthal angle $\phi$ over time while switching occurs, i.e. during the time $\theta_2$ changes from 175$^{\circ}$ to 5$^{\circ}$. (Reprinted with permission from Ref.~\phdref. Copyright 2012, Kuntal Roy.)}
\end{figurehere}

\section{\label{sec:conclusions}Summary and Outlook}

We have theoretically investigated electric-field induced switching of magnetization in a magnetostrictive nanomagnet strain-coupled with a piezoelectric layer in a multiferroic composite structure. The performance metrics of switching like sub-nanosecond switching delay and $\sim$1 attojoule energy dissipation are very promising for application purposes, which can extend the lifeline of conventional electronics. Also we have performed simulations for an array of such multiferroic devices to demonstrate that unidirectional signal propagation is possible in sub-nanosecond switching delay, which is particularly important for general-purpose computing. In the same way, logic gates and building blocks for circuits using multiferroic straintronic devices can be realized. Interestingly, the analysis of switching dynamics shows both qualitatively and quantitatively that steady-state analysis is highly inaccurate particularly in the context of switching process and determining switching delay. The out-of-plane excursion of magnetization highly facilitates the switching process and it can reduce the switching delay by a couple of orders in magnitude to sub-nanosecond. So multiferroic straintronic devices are intriguing in respect to both applied physics and basic physics of binary switching. Utilizing multiferroic composites for the purposes of room-temperature computing can be so energy-efficient that it can be powered solely by energy harvested from the environment. Therefore, they are ideal for medically implanted devices which draw energy solely from the patient's body movements, or even energy radiated by 3G networks and television stations. Moreover, the basic building blocks are non-volatile permitting instant turn-on computer. This is an unprecedented opportunity in ultra-low-energy computing and can perpetuate Moore's law to beyond the year 2020. Successful experimental implementation will pave the way for our future nanoelectronics.

\nonumsection{Acknowledgments} \noindent I acknowledge discussions with Jayasimha Atulasimha, Supriyo Bandyopadhyay, Supriyo Datta, and Avik Ghosh.


\end{multicols}
\end{document}